\documentclass[12pt]{article}

\usepackage{amssymb}
\usepackage{amsmath,amsthm, mathtools,commath}
\usepackage{graphics, color}
\usepackage{graphicx}
\usepackage{epsfig}
\usepackage{booktabs, arydshln}
\usepackage{caption}

\newcommand{\blind}{1}

\usepackage[normalem]{ulem}
\newcommand{\bomega}{\bm{\omega}}
\usepackage[round]{natbib}

\newcommand{\biblist}{\begin{list}{}
{\listparindent 0.0cm \leftmargin 0.50cm \itemindent -0.50 cm
\labelwidth 0 cm \labelsep 0.50 cm
\usecounter{list}}\clubpenalty4000\widowpenalty4000}
\newcommand{\ebiblist}{\end{list}}

\usepackage{comment} 
\usepackage{latexsym}
\usepackage{amsmath, amssymb, amsfonts, amsthm, bm}
\usepackage{graphicx}
\usepackage{mathrsfs}
\usepackage{algorithm, algpseudocode}

\newtheorem{theorem}{Theorem}
\newtheorem{corollary}{Corollary}

\newtheorem{remark}{Remark}

\newtheorem{lemma}{Lemma}
\newtheorem{condition}{Condition}

\usepackage{amsmath,amssymb,amsthm}
\usepackage{booktabs}
\usepackage{pgfplots}
\pgfplotsset{compat=1.18}

\newcommand{\tDG}{\widetilde D_G}
\newcommand{\DG}{D_G}
\newcommand{\wzero}{\bomega^{(0)}}

\DeclareMathOperator*{\argmin}{arg\,min}
\newcommand{\bX}{\mathbf{X}}

\newcommand{\bx}{\bm x}

\newcommand{\bz}{\bm z}
\newcommand{\bu}{\mathbf{u}}

\newcommand{\bbE}{\mathbb{E}}
\newcommand{\bbV}{{\mathbb{V}\rm{ar}}}
\newcommand{\bbP}{\mathbb{P}}

\newcommand{\bbeta}{\bm \beta}
\newcommand{\blambda}{\bm \lambda}
\newcommand{\bgamma}{\bm \gamma}
\newcommand{\bbx}{\mathbf{b}}
\newcommand{\bphi}{\bm \phi}
\def\T{\top}

\usepackage{multirow}
\usepackage{subfigure}

\usepackage{xr}
\numberwithin{equation}{section}

\allowdisplaybreaks[4]

\begin{document}
\def\spacingset#1{\renewcommand{\baselinestretch}%
{#1}\small\normalsize} 
\spacingset{1}

\if1\blind
{
  \title{\bf 
  A General Approach for Calibration Weighting under Missing at Random  }
  \author{Yonghyun Kwon \\
  Department of Mathematics, Korea Military Academy, Seoul, Republic of Korea \\
  and \\
   Jae Kwang Kim \\
  Department of Statistics, Iowa State University, Iowa, USA\\
  and \\
  Yumou Qiu \\
  School of Mathematical Sciences, Peking University, Beijing, China  
  }
\date{}
\maketitle
} \fi

\if0\blind
{
 \bigskip
 \bigskip
 \bigskip
 \begin{center}
   {\LARGE\bf 
   A General Approach for Calibration Weighting under Missing at Random}
\end{center}
 \medskip
} \fi

\bigskip

\begin{abstract}
We propose a unified class of calibration weighting methods based on weighted \emph{generalized entropy} to handle missing at random (MAR) data with improved stability and efficiency. The proposed generalized entropy calibration (GEC) formulates weight construction as a convex optimization program that unifies entropy-based approaches and generalized regression weighting. Double robustness is achieved by augmenting standard covariate balancing with a \emph{debiasing} constraint tied to the propensity score model and a \emph{Neyman-orthogonal} constraint that removes first-order sensitivity to nuisance estimation. Selection of the weights on the entropy function can lead to the optimal calibration estimator under a correctly specified outcome regression model. The proposed GEC weighting ha  a nice geometric characterization: the GEC solution is the Bregman projection of the initial weights onto a constraint set, which yields a 
generalized Pythagorean identity and a nested decomposition that quantifies the incremental “distance” paid for additional constraints. We also develop a high-dimensional extension with \emph{soft calibration} and a projection calibration constraint that preserves doubly robust inference. Two simulation studies are presented to compare the performance of the proposed method with the existing methods. 
\end{abstract}

\noindent%
\vfill

\newpage
\spacingset{1.9} 

\section{Introduction} 
\label{ch5sec1}

Missing data pose a persistent challenge in applied statistics, affecting fields ranging from survey sampling to social science and medical research. Ignoring missing observations often leads to selection bias and invalid inference, motivating the development of methods that correct this bias effectively.
A classical strategy is inverse probability weighting (IPW), which reweights observed cases by the inverse of their estimated response propensities. When the response propensity (RP) model is correctly specified, IPW yields unbiased estimates. However, in practice, IPW can be unstable  when some estimated probabilities are small or when the model is misspecified. 

Alternatively, outcome regression (OR) or imputation methods predict missing outcomes from observed covariates. These estimators are typically more stable but can also be biased when the regression model is misspecified. The augmented inverse probability weighting (AIPW) estimator combines the two approaches, achieving double robustness: it remains consistent if either the RP or OR model is correctly specified \citep{robins1994estimation, tsiatis2006}.   Among these, augmented inverse probability weighting (AIPW) is particularly popular and enjoys semiparametric efficiency when both models are correctly specified. Nevertheless, the efficiency of AIPW deteriorates under model misspecification, and the resulting weights can be unstable or even take negative values. 

A complementary line of research pursues calibration weighting, which adjusts the sample weights so that weighted covariate moments match target population moments \citep{deville1992calibration}. 
This perspective links inverse weighting to moment matching and motivates methods such as entropy balancing \citep{hainmueller2012} and  empirical likelihood calibration  \citep{Qin2002,  han2013, chan2016globally,  han2019general, liu2023biased}.  
More recently, regularized calibration has been extended to  high-dimensional settings  \citep{tan2020regularized}.  
Despite these advances, existing calibration estimators typically handle only covariate balance and do not fully address how to incorporate the two models into calibration and other inferential issues  when nuisance models are estimated.


In this paper, we develop a unified framework, termed the  generalized entropy calibration (GEC), that extends traditional calibration weighting through the lens of generalized-entropy optimization. The proposed method formulates the construction of calibration weights as a convex optimization problem generated by a strictly convex entropy function $G(\cdot)$ with a weight function $q(\cdot)$. By augmenting the standard covariate-balancing constraint with two additional constraints, a debiasing constraint tied to the response-propensity model and a Neyman-orthogonal constraint that eliminates first-order sensitivity to nuisance estimation, the GEC yields weights that are simultaneously stable, nonnegative, and doubly robust. {The weight function $q(\cdot)$ on the entropy is included to improve the efficiency of the GEC estimator. A selection procedure for the weights is proposed, which can lead to the optimal estimator under a correctly specified OR model but a misspecified RP model.
This differs from semi-parametric optimality, which requires both the OR and RP models to be correctly specified, and design-based optimality under known response probabilities.}

Originally introduced by \cite{Kwon2025} in the context of survey sampling with a known sampling mechanism, the GEC framework is extended here to incorporate estimated propensity scores. Specifically, the RP model is used to obtain inverse probability weights, while the Neyman-orthogonality constraint accounts for estimation errors in these weights. As a result, the proposed estimator achieves valid inference when either the OR  model or the RP model is correctly specified, without the need for distinct variance formulas under the two cases. 

The GEC formulation also admits a clean geometric characterization. The calibrated weights are the Bregman projection of the initial weights onto a space defined by the augmented constraints. This view leads to an equality-form of the generalized Pythagorean identity and a nested decomposition that quantifies the incremental divergence incurred by adding additional constraints, thereby offering a transparent diagnostic for over-constraint and limited overlap of the propensity scores. 
Finally, we extend the framework to high-dimensional settings by introducing soft calibration and projection calibration, which achieves exact balance along the principal calibrated direction estimated from the data. The resulting high-dimensional GEC (GEC-HD) estimator retains doubly robust inference under standard sparsity conditions.

The paper is organized as follows. Section \ref{ch5sec2} introduces the problem formulation and the optimal AIPW estimator under the OR model, which motivates the augmented regression approach and calibration estimation.  
Section \ref{ch5sec3} develops GEC with debiasing and orthogonality constraints and its dual characterization. Section \ref{sec:theorem} establishes some statistical properties, including doubly robust inference and optimal weight selection under the OR model. Section \ref{sec:geo} provides the geometric interpretation of the proposed method. Section \ref{sec:selection-1} treats high-dimensional soft calibration, and Section \ref{ch5sec5} presents 
comprehensive simulations and an empirical study using the National Health and Nutrition Examination Survey (NHANES) data. Section \ref{ch5sec6} concludes this paper. All the technical proofs are relegated to the supplementary material (SM).

\section{Optimal AIPW and augmented regression}
\label{ch5sec2}

Suppose that there are $N$ independently and identically distributed (i.i.d.) realizations of $(\bX, Y, \delta )$, denoted by $\{ (\bx_{i},y_{i},\delta_{i}):i=1,\ldots ,N \} $, where $y_i$ is a study variable subject to missingness, $\bx_{i} = (x_{i1}, \ldots, x_{i p_0})^{\T}$ is a $p_0$-dimensional vector of observed covariates, and $\delta_{i}$ is the response  indicator associated with unit $i$. In particular, $\delta_{i} = 1$ if $y_{i}$ is observed and $\delta_{i} = 0$ otherwise. Thus, instead of observing $(\bx_i, y_i, \delta_i)$, we only observe $(\bx_i, \delta_i y_i, \delta_i)$ for $i=1, \ldots, N$. Let $n = \sum_{i = 1}^{N} \delta_i$ be the number of respondents. 
We assume that the missingness mechanism is missing at random (MAR) in the sense of \cite{rubin1976}. Thus, the study variable $Y$ is independent of the missingness indicator $\delta$ given $\bX$; namely, $\delta \perp Y \mid \bX$. We also assume that $P( \delta =1 \mid \mathbf{X}=\bx)>0$ for all $\bx$ in the support of $\mathbf{X}$. 

Our target is the population mean $\theta = \mathbb E(Y)$. Consider the outcome regression (OR) model 
\begin{equation}
 y_i = m(\bx_i)+ e_i , 
\label{eq:1}
\end{equation}
where the error $e_i$ satisfies $\bbE( e_i \mid \bx_i) = 0$ and $\bbV( e_i \mid \mathbf{x}_i ) = v(\bx_i)$. We assume $v(\bx_i) = \sigma^2 \tilde{v}_i$, bounded away from zero.  We further assume that $m(\bx)$ lies in the linear span \begin{equation}
 m(\bx) = b_1(\bx_i) \beta_1 + \ldots + b_p(\bx_i) \beta_p = \mathbf{b}_i^\top \bbeta,
 \label{mx}
 \end{equation} 
where $\mathbf{b}_i=(b_1(\bx_i),\dots,b_p(\bx_i))^\top$ denotes basis functions with $b_1(\bx) = 1$, $\bbeta = (\beta_1, \ldots, \beta_p)^{\T}$, and $p$ is at the same order of $p_0$.
Under these model assumptions, assuming that $\tilde{v}_i$ are known, the best linear unbiased estimator of $\mathbb{E}(Y)$ is given by the linear regression estimator
\begin{equation}
\hat{\theta}_{\rm opt} = \frac{1}{N} \sum_{i=1}^N \bbx_i^{\T} \hat{\bbeta}_{\rm GLS} \mbox{ \ for \ }
\hat{\bbeta}_{\rm GLS} = \bigg( \sum_{i =1}^N \delta_i  \bbx_i \bbx_i^{\T} \tilde{v}_i^{-1} \bigg)^{-1}\sum_{i=1}^N \delta_i   \bbx_i y_i \tilde{v}_i^{-1}.
\label{gls}
\end{equation}

To protect potential bias of the regression estimator due to model misspecification, 
we often posit a response propensity (RP) model
$\pi(\bx_i;\phi)=\mathbb P(\delta_i=1\mid \bx_i)$. 
Let $\hat\pi_i=\pi(\bx_i;\hat\bphi)$, where $\hat\bphi$ is a consistent estimate (e.g.\,MLE) of $\bphi$.
The augmented inverse probability weighting (AIPW) estimator \citep{robins1994estimation} stabilizes estimation by incorporating an OR adjustment:
\begin{equation}
 \hat{\theta}_{\rm AIPW} = \frac{1}{N} \sum_{i=1}^N \hat{m} (\bx_i) + \frac{1}{N} \sum_{i=1}^N \frac{\delta_i}{ \hat{\pi}_i} \{ y_i - \hat{m} (\bx_i) \}, 
 \label{augp}
 \end{equation}
where $\hat{m} (\bx_i) = \bbx_i^{\T} \hat{\bbeta}$ for an estimate $\hat{\bbeta}$. The first term is the prediction component and the second term provides bias correction via inverse weighting and is also called the ``rectifier'' by \cite{angelopous2023}. 

It is well known that $\hat\theta_{\mathrm{AIPW}}$ is \emph{doubly robust}: it is consistent if either the RP or the OR model is correct, and is semiparametrically efficient when both models are correct \citep{robins1994estimation}. When the RP model is misspecified, however, efficiency depends on the choice of the estimated regression coefficient $\hat\bbeta$.

\subsection{Optimal AIPW estimator}
\label{sec:optimal-aipw}

To improve efficiency under a correct OR model but a possibly misspecified RP model, 
following the idea of \cite{magee98}, we consider the class of unbiased estimating equations
\begin{equation}
\sum_{i=1}^N \delta_i ( y_i - \mathbf{b}_i^\top \bbeta ) \mathbf{b}_i  q(\bx_i) = \mathbf{0}
 \label{qee}
\end{equation}
with scalar weight function $q(\cdot)$. Let $\hat\bbeta_q$ denote the solution. The AIPW estimators with the regression coefficient $\hat\bbeta_q$ form a family indexed by $q(\cdot)$:
\begin{equation}
\bigg\{\hat\theta_{\mathrm{AIPW},q}
=  \frac{1}{N}\sum_{i=1}^N \frac{\delta_i}{\hat{\pi}_i} y_i+ \bigg( \frac{1}{N} \sum_{i=1}^N \mathbf{b}_i   - \frac{1}{N} \sum_{i=1}^N \frac{\delta_i}{\hat{\pi}_i} \mathbf{b}_i \bigg)^\top \hat{\bbeta}_q: q(\bx) \mbox{ is nonegative} \bigg\}
.
\label{eq:aipwq}
\end{equation}
Note that $\hat{\theta}_{{\rm AIPW}, q}$ is doubly robust regardless of the choice of the weight function $q(\bx)$, but its variance depends on $q( \cdot)$.

Under the OR model in 
(\ref{eq:1}) and (\ref{mx}), we can express 
$\hat{\theta}_{{\rm AIPW}, q}
= N^{-1} \sum_{i=1}^N \mathbf{b}_i^\top \bbeta + N^{-1} \sum_{i=1}^N (\delta_i e_i / \hat{\pi}_i) - \hat{\Delta}_b^\top ( \hat{\bbeta}_q - \bbeta )$, where 
$$ \hat{\Delta}_b = \frac{1}{N }  \sum_{i=1}^N \frac{\delta_i}{\hat{\pi}_i} \mathbf{b}_i 
- \frac{1}{N} \sum_{i=1}^N \mathbf{b}_i
. $$ 
If the RP model is also correct, then $\hat{\Delta}_b = o_p(1)$ and the term $\hat{\Delta}_b^\top ( \hat{\bbeta}_q - \bbeta )$ due to estimation of $\bbeta$ is asymptotically negligible. In this case, the choice of $q(\bx)$ does not make any difference asymptotically.  However, when the RP model is incorrect, the term $\hat{\Delta}_b^\top ( \hat{\bbeta}_q - \bbeta )$ is not negligible,  and the asymptotic variance of $\hat{\theta}_{{\rm AIPW}, q}$ depends on $q(\bx)$. This shows that the choice of the weight function $q(\bx)$ plays a role in the efficiency of the AIPW estimator.

Let $\widehat{\mathbf{M}}_q = \sum_{i=1}^N \delta_i \mathbf{b}_i \mathbf{b}_i^\top  q(\bx_i) / N$ and $ \mathbf{M}_q = \mathbb{E} \big\{ \delta q( \bX) \bbx \bbx^\top \big\}$, where $\bbx^\top = (b_1(\bX), \ldots, b_p(\bX))$. Using the expansion $\hat{\bbeta}_q - \bbeta  = (N \widehat{\mathbf{M}}_q)^{-1} \sum_{i=1}^N  \delta_i \mathbf{b}_i q(\bx_i) e_i = \{(N \mathbf{M}_q)^{-1} \sum_{i=1}^N  \delta_i \mathbf{b}_i q(\bx_i) e_i \} + o_p( N^{-1/2} )$ under standard regularity conditions, we can obtain 
\begin{equation}
\hat{\theta}_{{\rm AIPW}, q}
= \frac{1}{N} \sum_{i=1}^N \mathbf{b}_i^\top \bbeta + \frac{1}{N} \sum_{i=1}^N \delta_i \bigg\{ \hat{\pi}_i^{-1}  - 
\hat{\Delta}_b^\top \mathbf{M}_q^{-1} \mathbf{b}_i q(\bx_i) \bigg\} e_i + o_p( N^{-1/2}). 
\end{equation}
Therefore, the asymptotic variance of $\sqrt{N}(\hat{\theta}_{{\rm AIPW}, q} - \theta)$ is 
\begin{equation} 
\mathsf{AVar}\big( \sqrt{N}(\hat{\theta}_{{\rm AIPW}, q} - \theta) \big) = \mathbb{V} \{ \mathbb{E} ( Y \mid \bX ) \} + 
\mathbb{E} \bigg[
\delta \bigg\{
\frac{1}{\pi^*(\bX)} - \Delta_b^{* \top} \mathbf{M}_q^{-1}\,q(\bX)\, \bbx \bigg\}^{2} {v}(\bX) 
\bigg],
\label{avar}
\end{equation} 
where $\pi^*(\bX)$ is the probability limit of $\hat{\pi}(\bX)$ and $\Delta_b^* = N^{-1} \sum_{i=1}^N (\delta_i / \pi^{*}_i - 1) \mathbf{b}_i$.

The optimal function $q^*(\bx)$ is obtained by minimizing the second term of the asymptotic variance in (\ref{avar}). As  $v(\bx) = \sigma^2 \tilde{v}_i$, for $q(\bx)$ belonging to a parametric class $\mathcal{F} ( \kappa)= \{ q(\bx; \kappa); \kappa \in \mathbb{R}^m  \}$, we can write $q(\bx) = q(\bx; \kappa)$ and construct the empirical loss function 
\begin{equation}
\hat{L} (\kappa) = \frac{1}{N}\sum_{i=1}^N \delta_i \big\{ \hat{\pi}_i^{-1}  - \hat{\Delta}_b^\top  \widehat{\mathbf{M}}_q^{-1}(\kappa) q(\bx_i; \kappa)\, \mathbf{b}_i 
\big\}^{2} \tilde{v}_i, 
\label{emp_loss}
\end{equation}
where $\widehat{\mathbf{M}}_q( \kappa) = N^{-1} \sum_{i=1}^N \delta_i \mathbf{b}_i  \mathbf{b}_i^\top  q(\bx_i; \kappa)$. We find the minimizer $\hat{\kappa}^* = \argmin_{\kappa} \hat{L} (\kappa)$ in (\ref{emp_loss}) and compute the optimal weight function $q^*( \bx)= q( \bx; \hat{\kappa}^*)$. 
%
If $\tilde{v}_i$ is unknown, we may use 
$\hat{v} (\bx_i) = ( y_i - \mathbf{b}_i^\top  \hat{\bbeta}_{\rm OLS} )^2$ to replace $\tilde{v}_i$ in (\ref{emp_loss}), where $\hat{\bbeta}_{\rm OLS} = ( \sum_{i =1}^N \delta_i \mathbf{b}_i \mathbf{b}_i^\top)^{-1} \sum_{i =1}^N \delta_i \mathbf{b}_i y_i$. 
The resulting AIPW estimator using $q^*(\bx)$ is optimal in the sense that it minimizes the asymptotic variance of $\hat{\theta}_{{\rm AIPW}, q}$ among its class in (\ref{eq:aipwq}) with $q(\bx) = q(\bx; \kappa)$.

\subsection{Augmented regression and calibration estimator}

The AIPW estimator can be interpreted as a prediction estimator with an augmentation term that corrects bias from missingness. 
In fact, the bias correction step can be inherently built in the predictor $\hat{m} (\bx_i)$ by augmented regression. Note that, 
for any predictor $\hat m(\bx_i)$ of $y_i$, if it satisfies the \emph{internal bias calibration (IBC)} condition \citep{firth1998}: 
\begin{equation}
\sum_{i =1}^N \frac{\delta_i }{\hat{\pi}_i} \left( y_i - \hat{m} (\bx_i) \right)=0,
\label{Ibc}
\end{equation} 
then the simple prediction estimator $N^{-1}\sum_{i = 1}^{N} \hat{m} (\bx_i)$ is doubly robust, since (\ref{Ibc}) implies 
\begin{eqnarray*} 
\frac{1}{N} \sum_{i=1}^N \hat{m} (\bx_i) = \frac{1}{N} \sum_{i=1}^N \bigg\{ \hat{m} (\bx_i) + \frac{\delta_i}{\hat{\pi}_i} ( y_i - \hat{m} (\bx_i) ) \bigg\},
\end{eqnarray*} 
which is consistent to $\theta$ if either the OR or RP model is correct. 

Now, we introduce the augmented regression approach to achieving the IBC condition in (\ref{Ibc}) and its doubly robust property. Recall that $q_i = q(\bx_i)$ is the weight for computing $\hat{\bbeta}_q$ in (\ref{qee}). The following lemma presents a sufficient condition for (\ref{Ibc}). 

\begin{lemma}\label{lm:1}
For $\hat{y}_i = \hat{m} (\bx_i) = \bbx_i^\top  \hat{\bbeta}_{q}$, if $( q_i  \hat \pi_i )^{-1}$ lies in the columns space of $\bbx_i$ for all observations with $\delta_i = 1$, then the IBC condition in (\ref{Ibc}) holds. 
\end{lemma}

From Lemma \ref{lm:1}, to satisfy (\ref{Ibc}), we can include $( \hat{\pi}_i {q}_i )^{-1} $ as an additional covariate in the augmented regression of $y_i$ on $\bbx_i$ and $( \hat{\pi}_i q_i)^{-1}$ to get $\hat{y}_i = \tilde{\bz}_i^{\T} \hat{\bgamma}_{q}$, which leads to the augmented prediction estimator $\hat{\theta}_{\rm AP} = N^{-1} \sum_{i = 1}^{N} \tilde{\bz}_i^{\T} \hat{\bgamma}_{q}$, where $\tilde{\bz}_i = (\bbx_i^{\T}, 1/( q_i \hat{\pi}_i))^{\T}$ and 
\begin{equation}
\hat{\bgamma}_q = \bigg( \sum_{i =1}^N \delta_i  \tilde{\bz}_i \tilde{\bz}_i^{\T} q_i \bigg)^{-1} \sum_{i=1}^N \delta_i \tilde{\bz}_i y_i q_i.
\label{gls2}
\end{equation}
The predictor $\hat{y}_i = \tilde{\bz}_i^{\T} \hat{\bgamma}_{q}$ using the augmented covariates $\tilde{\bz}_i$ satisfies the IBC condition in (\ref{Ibc}) by construction. Therefore, the augmented prediction estimator $\hat{\theta}_{\rm AP}$ is doubly robust. 
Similar as the AIPW estimator $\hat{\theta}_{{\rm AIPW}, q}$, the weights $\{q_i\}$ would not affect the double robustness of $\hat{\theta}_{\rm AP}$, but selection of $\{q_i\}$ could improve its efficiency.

This augmented-regression prediction estimator can equivalently be formulated as a \emph{calibration estimator} with an augmented balancing constraint.  To achieve unbiasedness under the outcome regression model in (\ref{eq:1}) with (\ref{mx}), the weights $\bomega$ need to satisfy
\begin{equation}
\sum_{i=1}^N \delta_i \omega_i \bbx_i = \sum_{i=1}^N \bbx_i.
\label{Calib}
\end{equation} 
To make the calibration estimator satisfy the double robustness, we obtain the calibration weights by the augmented constraint optimization:  
\begin{eqnarray}
& \tilde{\bomega} = \argmin_{\omega_i} \sum_{i=1}^N \delta_i \omega_i^2 q_i^{-1} & \mbox{ subject to (\ref{Calib}) and \ } \label{Wel0} \\
& & \ \sum_{i =1}^N \delta_i  \omega_i (q_i \hat{\pi}_i)^{-1} = \sum_{i=1}^N (q_i \hat{\pi}_i)^{-1}, \label{Wel0-IBC}
\end{eqnarray}
where $\tilde{\bomega} = (\tilde{\omega}_i: \delta_i =1)$. 
By the Lagrange multiplier method, the solution to (\ref{Wel0})--(\ref{Wel0-IBC}) 
is $\tilde{\omega}_i = \big( \sum_{i=1}^N \tilde{\bz} \big)^{\T} \big( \sum_{i=1}^N \delta_i \tilde{\bz}_i \tilde{\bz}_i^{\T} q_i \big)^{-1} \tilde{\bz}_i q_i$, which implies
the calibration estimator satisfies
\begin{equation}
\hat{\theta}_{\tilde{\bomega}} = \frac{1}{N} \sum_{i=1}^N \delta_i \tilde{\omega}_i y_i = \frac{1}{N} \sum_{i=1}^N \tilde{\bz}_i^{\T} \hat{\bgamma}_{q}
 = \frac{1}{N} \sum_{i=1}^N \bigg\{ \tilde{\bz}_i^{\T} \hat{\bgamma}_{q} + \frac{\delta_i}{\hat{\pi}_i} ( y_i - \tilde{\bz}_i^{\T} \hat{\bgamma}_{q} ) \bigg\}.
\label{opt3}
\end{equation}
Therefore, the calibration estimator $\hat{\theta}_{\tilde{\bomega}}$ with the additional constraint in (\ref{Wel0-IBC}) is the same as the augmented prediction estimator $\hat{\theta}_{\rm AP}$, and hence, it is doubly robust.

Although the calibration estimator that satisfies the IBC condition is doubly robust in estimation, the uncertainty associated with the estimated propensity score $\hat{\pi}_i = \pi( \bx_i; \hat{\bm \phi})$  is not reflected in the calibration procedure, making the resulting inference complicated. One remedy is to include an additional estimating equation to reflect the uncertainty of $\hat{\bphi}$ \citep{cao09}. This technique is  related to the so-called Neyman orthogonalization \citep{chernozhukov2018double} with respect to the nuisance parameter $\bphi$ in the RP model. How to achieve the Neyman orthogonality in the context of weight calibration has not been addressed in the literature. 
The optimal calibration estimator with respect to the choice of $\{q_i\}$ in (\ref{Wel0}) under a correctly specified OR model has not been studied either. 
In the following section, we propose a generalized entropy weighting method to tackle those problems and develop a doubly robust inference procedure.


\section{Generalized entropy calibration}
\label{ch5sec3}
To obtain non-negative weights and achieve other desirable properties, we develop a unified approach using generalized entropy of \cite{newey2004higher}. Let $G: \mathcal V \to \mathbb R$ be a prespecified function that is strictly convex and twice-continuously differentiable. The domain of $G$ is an open interval $\mathcal V = (\nu_1, \nu_2)$ in $\mathbb R$, where $\nu_1 > 0$ and $\nu_2$ is allowed to be $\infty$. Let $\pi^{\prime}(\bx_i, \bphi) = \partial \pi(\bx_i, \bphi) / \partial \bphi$ and $\pi^{\prime \prime}(\bx_i, \bphi) = \partial^2 \pi(\bx_i, \bphi) / \partial \bphi \partial \bphi^{\T}$ be the first and second order derivatives of the RP model with respect to $\bphi$, respectively.
Once $\hat {\bm \phi}$ and $\hat{\pi}_i = \pi(\bx_i, \hat{\bphi})$ are obtained,
the proposed calibration weighting can be formulated as the constraint optimization problem: 
\begin{eqnarray}
\hat{\bm \omega} &=& \argmin_{\omega_i \in \mathcal V}\sum_{i=1}^N \delta_i G( \omega_i) q_i^{-1} , 
\mbox{ \ subject to (\ref{Calib})}, \label{Wel} \\
&& \sum_{i =1}^N \delta_i \omega_i g(\hat{\pi}_i^{-1}) q_i^{-1}   = \sum_{i=1}^N g(\hat{\pi}_i^{-1}) q_i^{-1}  \mbox{ \ and } \label{con4} \\
&& \sum_{i=1}^N \delta_i \omega_i
\big( \partial_\phi \hat{g}_i \big)  q_i^{-1}  = \sum_{i=1}^N 
\big( \partial_\phi \hat{g}_i \big)  q_i^{-1}  
\label{con-phi} 
\end{eqnarray}
where $\hat{\bm \omega} = (\hat{\omega}_i: \delta_i = 1)$, $\partial_\phi \hat{g}_i = - g^{\prime}(\hat{\pi}_i^{-1}) \hat{\pi}_i^{-2}   \pi^{\prime}(\bx_i, \hat{\bphi})
$ and
$g(\omega) = d G(\omega)/ d \omega$ and $g^{\prime}(\omega) = d g(\omega)/ d \omega$ are the first-order and second-order derivatives of $G(\omega)$, respectively. The weight $q_i^{-1}$ in (\ref{Wel}) is treated as a known function of $\bx_i$. 
In Section 5, a data-driven method of finding an optimal $q_i$ will be discussed. 
The covariate balancing constraint in (\ref{Calib}) is associated with the OR model in \eqref{eq:1}. The constraint in (\ref{con4}) incorporates the RP model to achieve double robustness estimation while the constraint in (\ref{con-phi}) makes the resulting calibration estimator Neyman-orthogonal to the nuisance parameter ${\bphi}$ of the working RP model. 
We call (\ref{con4}) and (\ref{con-phi}) as the debiasing calibration constraint and the orthogonal calibration constraint for the entropy function $G(\omega)$, respectively.

Note that the regression weighting in (\ref{Wel0}) is a special case with $G(\omega_i) = \omega_i^2$, but without the orthogonal calibration constraint in (\ref{con-phi}). \cite{hainmueller2012} presented the exponential entropy $G(\omega) = \omega \log \omega$ and \cite{imai2014} suggested the empirical likelihood entropy $G(\omega) = \log \omega$, but their methods did not reflect the heterogeneous variance or consider the additional balancing constraints in \eqref{con4} and (\ref{con-phi}) to include the working RP model. For the special case of empirical likelihood where $G(\omega) = \log \omega$, the constraint in (\ref{con4}) becomes $\sum_{i =1}^N \delta_i \omega_i \hat{\pi}_i q_i^{-1}  = \sum_{i=1}^N \hat{\pi}_i q_i^{-1}$ which was considered in \cite{Han2014} for multiply robust estimation, and the constraint in (\ref{con-phi}) becomes $\sum_{i=1}^N \delta_i \omega_i q_i^{-1} \pi^{\prime}(\bx_i, \hat{\bphi}) = \sum_{i=1}^N q_i^{-1} \pi^{\prime}(\bx_i, \hat{\bphi})$ which was mentioned in \cite{chan2012uniform}. 
However, they did not consider the two constraints jointly, nor considered the heterogeneous variances $\{\tilde{v}_i\}$.

Let $\bz_i = \bz_i(\hat {\bm \phi}) = (\bbx_i^{\T}, \partial_\phi
 \hat{g}_i^{\T} q_i^{-1}, g(\hat{\pi}_i^{-1}) q_i^{-1})^{\T} \in \mathbb{R}^{p + p_0 + 1}$. Under suitable conditions, the strong duality for the constraint optimization problem in (\ref{Wel}) holds, and we can obtain the calibration weights $\hat{\bm \omega}$ in (\ref{Wel}) by the Lagrange multiplier method, which is the solution to the min-max problem $\min_{\bm \lambda}\max_{\bomega} Q(\bomega,\blambda)$, where $\bm \lambda = (\lambda_1, \ldots, \lambda_{p + p_0 + 1})^{\T}$ is the Lagrange multiplier, and 
\begin{align}
\label{qq} 
    Q(\bomega,\blambda) &= - \sum_{i=1}^N \delta_i G(\omega_i) q_i^{-1}  + \blambda^\top \bigg(\sum_{i=1}^N \delta_i \omega_i \bz_i  -\sum_{i=1}^N  \bz_i \bigg).
\end{align} 
Note that $\partial Q(\bomega,\blambda)/ \partial \omega_i = - g(\omega_i) q_i^{-1}+ \blambda^\top\bz_i  = 0$ for $\delta_i =1$, which leads to
  \begin{align}\label{Fwgt}
      \omega_i (\blambda, \hat{\bphi}) &= g^{-1}(\blambda^\top \bz_i q_i),
  \end{align}
  where $g^{-1}(\cdot)$ is the inverse function of $g(\cdot)$. 
By plugging (\ref{Fwgt}) into (\ref{qq}), we obtain 
\begin{eqnarray}
Q( \bomega(\blambda), \blambda) 
&=&  - \sum_{i=1}^N \delta_i G\{  g^{-1} ( \blambda^\top 
\bz_i q_i )\} q_i^{-1}+  \sum_{i=1}^N \delta_i g^{-1} ( \blambda^\top 
\bz_i q_i ) ( \blambda^\top 
\bz_i )  -    \sum_{i=1}^N ( \blambda^\top 
\bz_i ) 
\notag \\
&=&  \sum_{i=1}^N \delta_i q_i^{-1}  F( \blambda^\top \bz_i q_i) -  \sum_{i=1}^N ( \blambda^\top \bz_i),
\label{rho} 
\end{eqnarray}
where $F(\nu) = - G\{ g^{-1} ( \nu)\} + g^{-1} ( \nu) \nu$ is the convex conjugate function of $G$.  Therefore,  we can obtain $\hat{\blambda} = (\hat{\lambda}_1, \ldots, \hat{\lambda}_{p + p_0 + 1})^{\T}$ by  
\begin{equation}
\hat{\blambda} = \argmin_{\blambda} \hat{\rho} ( \blambda), \label{dual}
\end{equation}
where $\hat{\rho} ( \blambda) = N^{-1}Q( \bomega(\blambda), \blambda)  =  N^{-1} \big\{ \sum_{i = 1}^N \delta_i q_i^{-1} F\del{\bm \lambda^{\T} \bm z_i q_i} - \bm \lambda^{\T} \sum_{i = 1}^N \bm z_i \big\}$ is also a convex function. 
Since $\nabla \hat{\rho} (\blambda) =0$ gives the calibration equation, $\hat{\rho}( \blambda)$ is called the calibration generating function. 
Once $\hat{\blambda}$ is obtained from (\ref{dual}),  the calibration weights can be obtained by plugging  $\hat{\bm \lambda}$  to \eqref{Fwgt}. Examples of generalized entropies and their debiasing calibration constraints can be found in Table \ref{tab1}. 

\begin{table}[!t]
\centering
\resizebox{\columnwidth}{!}{%
\begin{tabular}{ccccc}
\hline\hline
Entropy                                             & $G(\omega)$     & $g_i = g\del{\pi_i^{-1}}$     & $1 / g^{\prime}(\pi_i^{-1})$      & Domain $\mathcal V$ \\ \hline
Empirical likelihood                                & $-\log \omega$              & $- \pi_i$         & $\pi_i^{-2}$            & $(0, \infty)$       \\
Exponential tilting                                 & $\omega\log (\omega) - \omega$ & $-\log \pi_i$         & $\pi_i^{-1}$                & $(0, \infty)$       \\
Contrast entropy                                       & $(\omega - 1)\log(\omega - 1) - \omega \log(\omega)$    & $\log(1 - \pi_i)$ & $\pi_i^{-2} - \pi_i^{-1}$  & $(1, \infty)$       \\
Hellinger distance                                  & $-4\sqrt{\omega}$            & $-2\pi_i^{1/2}$       & $\pi_i^{-3 / 2}$        & $(0, \infty)$       \\
Log-log & $-\log \del{\log \omega}$ &  $\pi_i (\log \pi_i)^{-1}$ & $ g_i^{-2}\{1 - \log \pi_i\}^{-1}$ & $(1, \infty)$ \\ 
Inverse  & $1 / (2\omega)$   & $-\pi_i^2 / 2$  & $\pi_i^{-3}$  & $(0, \infty)$ \\
R\'enyi entropy & $\alpha^{-1}(\alpha + 1)^{-1}\omega^{\alpha + 1}$   & $\alpha^{-1} \pi_i^{-\alpha}$ & $ \pi_i^{\alpha - 1}$ & $(0, \infty)$ \\ \hline\hline
\end{tabular}
}
\caption{Examples of generalized entropies with the corresponding $G(\omega)$, the calibration covariates $g_i = g(\pi_i^{-1})$, and the regression weight $1 / g^{\prime}(\pi_i^{-1})$ in (\ref{Gammahat}), where $g^{\prime}(\cdot)$ denotes the first-order derivative of $g(\cdot)$. R\'enyi entropy requires $\alpha \neq 0, -1$.} 
\label{tab1}
\end{table}

Let $\hat{\omega}_i = {\omega}_i(\hat{\bm \lambda}, \hat{\bm \phi})$ denote the solution of the constraint optimization problem in (\ref{Wel}), which emphasizes its dependence on $\hat{\bm \lambda}$ in (\ref{Fwgt}) and $\hat{\bm \phi}$ in the estimated RP model $\hat{\pi}_i = \pi( \bx_i; \hat{\bm \phi})$. The proposed generalized entropy calibration (GEC) estimator of $\theta = \bbE(Y)$ is
\begin{equation}
    \hat{\theta}_{\rm GEC} = N^{-1} \sum_{i=1}^N \delta_i \hat{\omega}_i y_i = N^{-1} \sum_{i=1}^N \delta_i \hat{\omega}_i(\hat{\bm \lambda}, \hat{\bm \phi}) y_i.
\label{eq:GEC}
\end{equation}
In the following, we explain the rationale of the doubly robust inference property of the GEC estimator and its asymptotic expansion for statistical inference. 

Let $f(\nu) = d F(\nu)/ d \nu$ be the derivative of $F(\nu)$ and $z_{ij}(\hat{\bphi})$ be $j$th component of $\bz_i(\hat{\bphi})$. Then, $f(\nu) = g^{-1}(\nu)$ and $\omega_i (\hat{\blambda}, \hat{\bphi}) = f(\hat{\blambda}^\top\bz_i(\hat{\bphi})q_i)$.
Note that $\partial {\omega}_i(\hat{\bm \lambda}, \hat{\bm \phi}) / \partial \blambda = f^{\prime} ( \hat{\bm \lambda}^{\T} \bm z_i(\hat{\bm \phi})q_i ) \bm z_i(\hat{\bm \phi}) q_i = \{g^{\prime}(\hat{\omega}_i)q_i^{-1}\}^{-1} \bm z_i$ and $\partial {\omega}_i(\hat{\bm \lambda}, \hat{\bm \phi}) / \partial \bphi = - \{g^{\prime}(\hat{\omega}_i)\}^{-1} \{ \sum_{j = p + 1}^{p + q} \hat{\lambda}_j z_{ij}^{\prime}(\hat{\bphi}) - \hat{\lambda}_{p + p_0 + 1} g^{\prime}(\hat{\pi}_i^{-1}) \hat{\pi}_i^{-2} q_i^{-1} \pi^{\prime}(\bx_i, \hat{\bphi}) \} q_i$, where $z_{ij}^{\prime}(\hat{\bphi}) = \partial z_{ij}(\hat{\bphi}) / \partial \bphi$.
Let
$$ \hat{\theta}_{\rm GEC} ( \hat{\bm \lambda}, \hat{\bm \phi}, \bgamma) = \frac{1}{N} \sum_{i=1}^N \delta_i  \omega_i(\hat{\blambda}, \hat{\bm \phi}) y_i + \frac{1}{N} \bigg( \sum_{i=1}^N \bz_i - \sum_{i=1}^N \delta_i\omega_i(\hat{\blambda}, \hat{\bm \phi}) \bz_i \bigg)^{\T} \bgamma. 
$$
Due to the balancing constraints in (\ref{Calib}) and (\ref{con4}), we have that $\hat{\theta}_{\rm GEC} = \hat{\theta}_{\rm GEC} ( \hat{\blambda}, \hat{\bphi}, \bgamma)$ for all $\bgamma$. Furthermore, if we choose $\hat{\bgamma}$ as the solution to 
\begin{equation}
 \sum_{i=1}^N \delta_i \{g^{\prime}(\hat{\omega}_i) q_i^{-1}\}^{-1} ( y_i - \bz_i^{\T} \bgamma )\bm z_i = \mathbf{0}, 
 \label{eq:13}
\end{equation}
we have $\partial \hat{\theta}_{\rm GEC} ( \hat{\bm \lambda}, \hat{\bm \phi}, \hat{\bgamma}) / \partial \blambda = \mathbf{0}$, 
meaning the effect of estimating $\blambda$ on $\hat{\theta}_{\rm GEC}$ can be safely ignored \citep{randles1982}. Meanwhile, 
$$\partial \hat{\theta}_{\rm GEC} ( \hat{\bm \lambda}, \hat{\bm \phi}, \bgamma) / \partial \bphi = \frac{1}{N} \sum_{i = 1}^{N} \delta_i (y_i - \bz_i^{\T}\bgamma) \partial {\omega}_i(\hat{\bm \lambda}, \hat{\bm \phi}) / \partial \bphi + \frac{1}{N} \sum_{i = 1}^{N} (1 - \delta_i \omega_i(\hat{\bm \lambda}, \hat{\bm \phi})) \partial (\bm z_i^{\T} \bgamma) / \partial \bphi.$$
If the OR model in (\ref{eq:1}) and (\ref{mx}) is correctly specified such that $m(\bx_i) = \bbx_i^{\T} \bbeta_0$ for a $\bbeta_0 \in \mathbb{R}^p$, we have $\hat{\bgamma} \overset{p}{\to} (\bbeta_0^{\T}, \bm 0_{p_0 + 1}^{\T})^{\T}$, where $\bm 0_{p_0 + 1}$ denotes a $(p_0 + 1)$-dimensional vector of zero. Under this case, $\partial \hat{\theta}_{\rm GEC} ( \hat{\bm \lambda}, \hat{\bm \phi}, \hat{\bgamma}) / \partial \bphi \overset{p}{\to} \bm 0$. If the RP model $\pi(\bx_i, \bphi_0)$ is correctly specified and $\hat{\bphi} \overset{p}{\to} \bphi_0$, it can be shown that $\hat{\lambda}_j \overset{p}{\to} 0$ for $j = 1, \ldots, p + p_0$, which implies that $\partial \hat{\theta}_{\rm GEC} ( \hat{\bm \lambda}, \hat{\bm \phi}, \hat{\bgamma}) / \partial \bphi \approx - N^{-1} \hat{\lambda}_{p + p_0 + 1} \sum_{i = 1}^{N} \delta_i \{g^{\prime}(\hat{\omega}_i) q_i^{-1}\}^{-1} (y_i - \bz_i^{\T} \hat{\bgamma}) g^{\prime}(\hat{\pi}_i^{-1}) \hat{\pi}_i^{-2} q_i^{-1} \pi^{\prime}(\bx_i, \hat{\bphi}) = \bm 0$ as $\bm z_i$ includes the additional covariates $\partial_\phi \hat{g}_i q_i^{-1}$. Thus, $\partial \hat{\theta}_{\rm GEC} ( \hat{\bm \lambda}, \hat{\bm \phi}, \hat{\bgamma}) / \partial \bphi \overset{p}{\to} \bm 0$ under either correct OR or correct RP model, which implies the Neyman orthogonality property \citep{chernozhukov2018double} of the proposed calibration weighted estimator $\hat{\theta}_{\rm GEC}$ and the estimation error in $\hat{\pi}_i$ can be safely ignored. 

Note that the solution to (\ref{eq:13}) is 
\begin{equation}
\hat{\bm \gamma} = \bigg( 
\sum_{i=1}^N \frac{\delta_i q_i \bz_i \bz_i^{\T}}{g^{\prime}(\hat{\omega}_i)  }\bigg)^{-1} \sum_{i =1}^N \frac{\delta_i q_i \bz_i y_i} {g^{\prime}(\hat{\omega}_i) }. \label{Gammahat}
\end{equation}
Let $\blambda^*$, $\bm \phi^*$ and $\bgamma^*$ be the probability limits of $\hat{\blambda}$, $\hat {\bm \phi}$ and $\hat{\bgamma}$ under either a correctly specified OR or RP model, respectively. Let $\pi_i^* = \pi(\bm x_i; \bm \phi^*)$, $\omega_i^* = \omega_i(\bm \lambda^*, \bm \phi^*) = f ( {\bm \lambda}^{* \T} \bm z_i^* q_i )$ and $\bz_i^* = \bz_i(\bm \phi^*) = (\bbx_i^{\T}, - g^{\prime}(1 / \pi_i^*) (\pi_i^*)^{-2} q_i^{-1} \pi^{\prime}(\bx_i, \bphi^*)^{\T}, g(1 / \pi_i^*) q_i^{-1})^{\T}$.
The above derivation implies the asymptotic expansion of the proposed GEC estimator $\hat{\theta}_{\rm GEC}$ as
\begin{equation}
\hat{\theta}_{\rm GEC} = \frac{1}{N} \sum_{i=1}^N \delta_i  \omega_i^* y_i + \frac{1}{N} \bigg\{ \sum_{i=1}^N (1 - \delta_i\omega_i^*) \bz_i^* \bigg\}^{\T} \bgamma^* + o_p(N^{-1/2})
\label{eq:asy-exp}\end{equation}
under either a correctly specified OR or RP model. This shows the doubly robust inference property of $\hat{\theta}_{\rm GEC}$ in the sense that its influence function is the same under either the OR or RP model. The following section provides the rigorous theoretical statement of (\ref{eq:asy-exp}).



\section{Statistical properties}\label{sec:theorem}

In this section, we provide the theoretical results for the proposed GEC estimator $\hat{\theta}_{\rm GEC}$. Recall that $\blambda^*$, $\bm \phi^*$ and $\bgamma^*$ are the probability limits of $\hat{\blambda}$, $\hat {\bm \phi}$ and $\hat{\bgamma}$, respectively, and $\pi_i^* = \pi(\bm x_i; \bm \phi^*)$, $\omega_i^* = f ( {\bm \lambda}^{* \T} \bm z_i^* q_i )$ and $\bz_i^* = \bz_i(\bm \phi^*)$. Their rigorous definitions are given in the SM. 
Let $\pi_0(\bx_i) = \bbP(\delta_i=1 \mid \bx_i)$ be the true propensity score. Recall that $\pi^{\prime}(\bx_i, \bphi)$ and $\pi^{\prime\prime}(\bx_i, \bphi)$ are the first and second order derivatives of $\pi(\bx_i, \bphi)$. Let $\|\cdot\|$ and $\|\cdot\|_{\rm F}$ denote the Euclidean norm for vectors and the Frobenius norm for matrices, respectively. We make the following conditions to facilitate our analysis. 

\begin{condition}\label{con:G}
The function $G(\omega): \mathcal{V} \to \mathbb{R}$ is strictly convex and continuously differentiable, with $G''(\omega) > 0$ for all $\omega \in \mathcal{V}$, where $\mathcal{V}$ is an open interval $\mathcal V  \subset (0, \infty)$.
\end{condition}

\begin{condition}\label{con:ps}
There exists a positive constant $c_0\in(0,1/2)$ such that the true propensity score $\pi_0(\bx_i)$ satisfies $c_0 \leq \pi_0(\bx_i) \leq 1-c_0$, $\{\pi_0(\bx_i)\}^{-1} \in \mathcal{V}$ and $c_0 \leq q_i \leq c_0^{-1}$ for $1\leq i \leq N$.
\end{condition}

\begin{condition}\label{con:RPmodel}
The RP model $\pi(\bx_i,\bphi)$ is (i) second-order continuously differentiable with respect to $\bphi$, $\mathbb{E}\{ \sup_{\|\bphi - \bphi^*\| \leq \epsilon}\|\pi'(\bx_i,\bphi)\|^{4 + \tilde{c}} \} < \infty$ and $\mathbb{E}\{ \sup_{\|\bphi - \bphi^*\| \leq \epsilon}\|\pi''(\bx_i,\bphi)\|_{\rm F}^{2 + \tilde{c}} \} < \infty$ for small positive constants $\epsilon$ and $\tilde{c}$; 
(ii) $c_0 \leq \pi(\bx_i,\bphi^*) \leq 1-c_0$ for $1 \leq i \leq N$ and $c_0 \in (0, 1/2)$; 
(iii) the estimate $\hat{\bphi}$ satisfies $\|\hat{\bphi} - \bphi^*\| = O_p(N^{-1/2})$.
(iv) Furthermore, if the RP model $\pi(\bx_i, \bphi)$ is misspecified, $c_0 \leq 1 / \omega_i^* \leq 1-c_0$ and $\omega_i^* \in \mathcal{V}$ for $1 \leq i \leq N$, where $\omega_i^* = f ( {\bm \lambda}^{* \T} \bm z_i^* q_i )$.
\end{condition}

\begin{condition}\label{con:nuisance}
The population values $\blambda^*$, $\bm \phi^*$ and $\bgamma^*$ of the nuisance parameters, defined in (S.1)--(S.3) in the SM, exist and are unique. 
\end{condition}

\begin{condition}\label{con:cov}
(i) The covariates $\bm b_i$ satisfy $\mathbb{E}|b_j(\bx_i)|^{2 + \tilde{c}} \leq \infty$ for $j = 1, \ldots, p$.
(ii) The matrix $\bm \Sigma_{\bm z} = \mathbb{E}(\bm z_i^* \bm z_i^{* \T})$ is positive definite. 
\end{condition}

Condition \ref{con:G} regulates the smoothness and convexity of the entropy function $G(\omega)$. All the entropy functions listed in Table \ref{tab1} satisfy this condition. Condition \ref{con:ps} makes the strongly overlapping condition on the true propensity scores, which is common in missing data and causal inference literature. It also assumes the weights $\{q_i\}$ used for the generalized entropy optimization in (\ref{Wel}) are bounded. Condition \ref{con:RPmodel} regulates the smoothness of the working RP model, and the moment of its first and second-order derivatives with respect to $\bphi$. Those conditions are necessary to control the difference between $\pi(\bx_i, \hat{\bphi})$ and $\pi(\bx_i, \bphi^*)$ in the analysis. The proposed method does not rely on a particular estimator of the RP model. We only need the estimate of $\bphi$ is $\sqrt{N}$-consistent. The assumptions on the limits of the nuisance parameters of the working models in Conditions \ref{con:RPmodel} and \ref{con:nuisance} are standard for the case of misspecified RP and OR models. Similar conditions are made for doubly robust estimation in \cite{Tan2020Modelassited, ning2020robust}.
Note that $\bphi^* = \bphi_0$, $\lambda_j^* = 0$ for $j = 1, \ldots, p + p_0$ and $\lambda_{p + p_0 + 1}^* = 1$ if the RP model is correctly specified, and $\bgamma^* = (\bbeta_0^{\T}, \bm 0_{p_0 + 1}^{\T})^{\T}$ if the OR model is correctly specified, where $\bm 0_{p_0 + 1}$ denotes a $(p_0 + 1)$-dimensional vector of zero.
Condition \ref{con:cov} is not restrictive, which only assumes the $2 + \tilde{c}$ moments of the covariates $\bm b_i$ exist for a smaller constant $\tilde{c}$, and the second moment of the augmented calibration functions $\bm z_i^*$ is positive definite.

The following lemma shows that the solution to the primal problem in (\ref{Wel}) exists and the solution $\hat{\blambda}$ to the corresponding dual problem in (\ref{dual}) converges to its probability limit $\blambda^*$ at the rate $O_p(N^{-1/2})$. 

\begin{lemma}
Under Conditions \ref{con:G}--\ref{con:cov}, the solution $\hat {\bm \omega}$ to (\ref{Wel}) exists and is unique with probability approaching to 1 as $N \to \infty$.
Furthermore, the solution $\hat{\blambda}$ to the corresponding dual problem in (\ref{dual}) satisfies $\|\hat{\blambda} - \blambda^*\| = O_p(N^{-1/2})$. 
\end{lemma}

Let $\bm \Gamma^* = \mathbb{E} \big[\pi_0(\bx_i) \{g^{\prime}(\omega_i^*) q_i^{-1}\}^{-1} \bz_i^* \bz_i^{* \T} \big]$ and $\bgamma^* = (\bm \Gamma^*)^{-1} \mathbb{E} \big[\pi_0(\bx_i) \{g^{\prime}(\omega_i^*) q_i^{-1}\}^{-1} \bz_i^* y_i \big]$. The following theorem establishes the asymptotic expansion of the proposed GEC estimator $ \hat{\theta}_{\rm GEC} = N^{-1} \sum_{i=1}^N \delta_i \hat{\omega}_i y_i$ in (\ref{eq:GEC}) under either a correctly specified RP or OR model. 

\begin{theorem}\label{thm:1}
Under Conditions \ref{con:G}--\ref{con:cov}, if either the RP model $\pi(\bx_i, \bphi)$ is correctly specified or the OR model $m(\bx_i) = \bbx_i^{\T} \bbeta_0$ in (\ref{eq:1}) and (\ref{mx}) is correctly specified, we have 
$$
\hat{\theta}_{\rm GEC} = \frac{1}{N} \sum_{i=1}^N \delta_i  \omega_i^* y_i + \frac{1}{N} \bigg\{ \sum_{i=1}^N (1 - \delta_i\omega_i^*) \bz_i^* \bigg\}^{\T} \bgamma^* + o_p(N^{-1/2}),
$$
and $\sqrt{N}(\hat{\theta}_{\rm GEC} - \theta) \overset{d}{\to} N(0, V_{\rm GEC})$ as $N \to \infty$, where $$V_{\rm GEC} = \bbV\{\delta_i  \omega_i^* y_i + (1 - \delta_i\omega_i^*) \bz_i^{* \T} \bgamma^*\}.$$
\end{theorem}

From Theorem \ref{thm:1}, the proposed GEC estimator is asymptotically normal with the same influence function under either a correct RP or OR model. The asymptotic variance $V_{\rm GEC}$ can be estimated by 
{
\begin{equation}
    \hat{V}_{\rm GEC} = \frac{1}{N-1} \sum_{i = 1}^{N} (\hat \eta_i  - \bar \eta)^2,
\label{eq:Vest}
\end{equation}
where $\hat \eta_i = \delta_i \hat{\omega}_i y_i + (1 - \delta_i \hat{\omega}_i) \bz_i^{\T} \hat{\bgamma}$, $\bar \eta = N^{-1}\sum_{i=1}^N \hat \eta_i$,
}
the generalized entropy weights $\{\hat{\omega}_i\}$ are the solution to the problem in (\ref{Wel}), $\bz_i = \bz_i(\hat {\bm \phi})$ and $\hat{\bm \gamma}$ is given in (\ref{Gammahat}). Due to the consistency of $\hat{\omega}_i$, $\hat {\bm \phi}$ and $\hat{\bm \gamma}$, it follows that $\hat{V}_{\rm GEC} \overset{p}{\to} V_{\rm GEC}$. Therefore, the $(1 - \alpha)$ confidence interval of $\theta$ can be constructed as 
\begin{equation}
    (\hat{\theta}_{\rm GEC} - z_{\alpha / 2}(\hat{V}_{\rm GEC} / N)^{1/2} , \hat{\theta}_{\rm GEC} + z_{\alpha / 2}(\hat{V}_{\rm GEC} / N)^{1/2}),
\label{eq:CI}
\end{equation}
where $z_{\alpha / 2}$ is the upper $\alpha / 2$ quantile of the standard normal distribution. 
Using the result in Theorem \ref{thm:1} and the consistency of the variance estimate $\hat{V}_{\rm GEC}$,  the proposed confidence interval in (\ref{eq:CI}) has the valid coverage asymptotically if either the RP or OR model is correctly specified. This shows the doubly robust inference property of the proposed estimator, achieved by carefully designing the debiasing and orthogonal calibration constraints in (\ref{con4}) and (\ref{con-phi}). 

Note that the inference procedure for the AIPW estimator $\hat{\theta}_{\rm AIPW}$ in (\ref{augp}) could depend on the influence function of the estimator $\hat{\bphi}$ as it is not Neyman orthogonal to the nuisance parameter $\bphi$ and has different variance estimators under the RP and OR models. 
However, we usually do not know which model is correct in practice. Therefore, the proposed estimator is more advantageous than the AIPW estimator in achieving doubly robust inference. Moreover, the GEC estimator could be more efficient than the AIPW estimator due to the additional covariates $
\partial_\phi \hat{g}_i 
q_i^{-1} $ and $g(\hat{\pi}_i^{-1}) q_i^{-1}$ used for balancing. See Remark 2 below. Compared to the existing doubly robust inference approach that uses specific calibration estimator for the working RP model \citep{tan2020regularized}, the proposed method provides a general calibration approach which do not have any restriction on the estimation of the RP model. 

The following two corollaries provide the specific influence functions of $\hat{\theta}_{\rm GEC}$ under the correctly specified OR and RP models, respectively, which is a direct application of Theorem \ref{thm:1} using the corresponding values of $\blambda^*$ and $\bgamma^*$ under each case.

\begin{corollary}\label{cy:1}
Under Conditions \ref{con:G}--\ref{con:cov}, if the OR model $m(\bx_i) = \bbx_i^{\T} \bbeta_0$ in (\ref{eq:1}) and (\ref{mx}) is correctly specified, we have 
$$
\hat{\theta}_{\rm GEC} = \frac{1}{N} \sum_{i=1}^N \bbx_i^\top {\bbeta}_0 + \sum_{i=1}^N \delta_i \omega_i^* e_i  + o_p(N^{-1/2}) , $$
where $e_i = y_i -\bbx_i^{\T} \bbeta_0$. Thus, 
$$V_{\rm GEC}  =  \bbV \{ m( \bX) \} + \bbE \{ \delta  {\omega^*}^2 v(\bX) \}, $$ 
where $\omega^* = f ( {\bm \lambda}^{* \T} \bm z(\bphi^*) q(\bX) )$ and $v(\bx)=\bbV( Y \mid \bx)$. \end{corollary}

\begin{remark}
By Corollary \ref{cy:1}, under the OR model, 
 the optimal choice of $q_i$  is obtained by minimizing the second term of $V_{\rm GEC}$: 
\begin{equation}
 \mathbb{E} \big[ \delta_i \{ f( q_i {\bm \lambda}^{* \T} \bm z_i^* ) \}^2 e_i^2 \big] .
 \label{eq:var2}
 \end{equation}
 Now, let $q_i = q( \bx_i; \kappa)$, which is parameterized by $\kappa$. We can obtain the optimal choice of $\kappa$ by minimizing 
$$ M( \kappa) = \sum_{i=1}^N \delta_i \big\{ f(q( \bx_i; \kappa) \bz_i^\top  \hat{\bm \lambda}_q) \big\}^2 \hat{e}_i^2 $$
with respect to $\kappa$, where $\hat{e}_i = y_i - \bz_i^\top \hat{\bm \gamma}_q$, $\hat{\bm \gamma}_q$ is defined in (\ref{Gammahat}), and $\hat{\bm \lambda}_q$ is obtained from (\ref{dual}). 
It is worth noting that the quantity in (\ref{eq:var2}) corresponds directly to the second term of (\ref{avar}), which governs the efficiency of the AIPW estimator. 
Specifically, for the square entropy $G(\omega) = \omega^2 / 2$, it can be shown that 
$\mathbb{E} \big[ \delta_i \{ f( q_i {\bm \lambda}^{* \T} \bm z_i^* ) \}^2 e_i^2 \big]$ has the same expression as the second term of (\ref{avar}).
This correspondence shows that the GEC optimization generalizes the variance-minimization principle in (\ref{avar}) within a unified convex optimization framework. 
\end{remark}

\begin{corollary}\label{cy:2}
Under Conditions \ref{con:G}--\ref{con:cov}, if the RP model $\pi(\bx_i, \bphi)$ is correctly specified, we have 
$$
\hat{\theta}_{\rm GEC} = \frac{1}{N} \sum_{i=1}^N \delta_i \pi_0^{-1}(\bx_i) y_i + \frac{1}{N} \sum_{i=1}^N \{1 - \delta_i\pi_0^{-1}(\bx_i)\} \bz_i^{* \T} \bgamma^* + o_p(N^{-1/2})
$$
and $$V_{\rm GEC} = \bbV(Y) + \bbE \big[ \{\pi_0^{-1}(\bX) -1\} \{Y - \bm z(\bm \phi_0)^{\T} \bm \gamma^* \}^2 \big], $$ where $\pi_0(\bx)=\pi(\bx; \phi_0)$ is the true response probability function. 
\end{corollary}

\begin{remark}\label{rm:2}
It is worth noticing that the main order of the variance of the AIPW estimator $\hat{\theta}_{\rm AIPW}$ in \eqref{augp} under the correct RP model is 
$\mathsf{AVar}\big( \sqrt{N}(\hat{\theta}_{\rm AIPW} - \theta) \big) = \bbV(Y) + \bbE \big[ \{ \pi_0^{-1}(\bX) - 1 \} \{ Y - \bX^{\T} \bbeta^* - \bm h(\bm \phi_0)^{\T} \bm \kappa^*\}^2 \big]$,
where $\bbeta^*$ is the probability limit of $\hat{\bbeta}$, $\bm h(\bm \phi)$ is the balancing functions used to estimate the RP model and ${\bm \kappa}^*$ is a coefficient vector.
If the OR model in (\ref{mx}) is also correct with $\bm b_i = \bx_i$, then $\bm z(\bm \phi_0)^{\T} \bm \gamma^* = \bX^{\T} \bbeta^* = \bX^{\T} \bbeta_0$ and $\bm \kappa^* = \bm 0$. In this case, the variances of $\hat{\theta}_{{\rm GEC}}$ and $\hat{\theta}_{\rm AIPW}$ are asymptotically equivalent. 
However, if the OR model is incorrect, $\bbV( \hat{\theta}_{{\rm GEC}})$ is likely to be smaller than $\bbV(\hat{\theta}_{\rm AIPW})$ as an additional covariate $g( \hat{\pi}_i^{-1} ) q_i^{-1}$ is included in the linear regression of $Y$, which can contribute to the prediction of $Y$. Therefore, the model-assisted calibration estimator using the augmented covariate $\bz_i$ for calibration could be more efficient than the classical AIPW estimator when the OR model in (\ref{eq:1}) and (\ref{mx}) is incorrect. 
\end{remark}


\section{Geometric interpretation}
\label{sec:geo}

We now present a geometric interpretation of the proposed GEC  method. For a given convex function $G(\cdot)$, define 
$$  D_G \big( \omega_i \parallel \omega_i^{(0)} \big) = G( \omega_i ) - G( \omega_i^{(0)} ) - g \big( \omega_i^{(0)} \big) \big( \omega_i - \omega_i^{(0)} \big)  
$$
to be the Bregman divergence of $\omega_i$ evaluated at $\omega_i^{(0)}$ using $G( \cdot)$ as a generator. The Bregman divergence represents the difference between $G(\omega_i)$ and its tangent line evaluated at $\omega_i^{(0)}$. Since $G( \cdot)$ is strictly convex, we can establish $D_G ( \omega_i \parallel \omega_i^{(0)} ) \ge 0$ with the equality at $\omega_i=\omega_i^{(0)}$. 

In our setup,  we may use $\omega_i^{(0)}=\hat{\pi}_i^{-1}$ as the initial weight and find the minimizer of 
\begin{equation}
 \sum_{i=1}^N \delta_i  D_G \big( \omega_i \parallel \omega_i^{(0)} \big) q_i^{-1} 
 \label{eq:6-1}
 \end{equation}
subject to some calibration constraints. If the calibration constraints include the debiasing constraint in (\ref{con4}), then we obtain $\sum_{i=1}^N \delta_i g(\omega_i^{(0)}) (\omega_i - \omega_i^{(0)}) q_i^{-1} = \sum_{i=1}^N (1 - \delta_i \omega_i^{(0)}) g(\omega_i^{(0)}) q_i^{-1} $, which leads to
\begin{eqnarray}
\sum_{i=1}^N \delta_i  D_G \big( \omega_i \parallel \omega_i^{(0)} \big) q_i^{-1} 
 =  \sum_{i =1}^N \delta_i G( \omega_i) q_i^{-1} + C_N, \nonumber
\end{eqnarray}
where $C_N$ is a constant free of $\omega_i$.
This result implies that the proposed calibration method described in Section \ref{ch5sec3} can be equivalently described as minimizing the total Bregman divergence in (\ref{eq:6-1}) subject to the same constraints. 

Using Bregman divergence, 
our calibration problem can be formulated as 
\begin{equation}
\hat{\bomega} = \mbox{arg} \min_{\omega \in \mathcal{C}} 
\sum_{i=1}^N \delta_i  D_G ( \omega_i \parallel \omega_i^{(0)} ) q_i^{-1} 
\label{eq:6-3}
\end{equation}
where 
$\mathcal{C}= \{ \bomega ; \sum_{i=1}^N \delta_i \omega_i \bz_i = \sum_{i=1}^N \bz_i \}
$ is the set of weights satisfying the calibration constraints on $\bz_i$.  
The solution (\ref{eq:6-3}) can be understood as the information projection of $\bomega^{(0)}$ onto the set 
$\mathcal{C}$ (m‑flat). By the same argument for obtaining (\ref{Fwgt}), the solution to the optimization problem in (\ref{eq:6-3}) can be expressed as 
\begin{align}\label{Fwgt2}
      \hat{\omega}_i  &= g^{-1}\big\{ g( \omega_i^{(0)}) +  \hat{\blambda}^\top \bz_i q_i \big\},
  \end{align}
where  
\begin{equation*}
\hat{\blambda} = \argmin_{\blambda} \bigg\{ \sum_{i = 1}^N \delta_i q_i^{-1} F\big(
g( \omega_i^{(0)}) + 
\bm \lambda^{\T} \bm z_i q_i \big) - \bm \lambda^{\T} \sum_{i = 1}^N \bm z_i \bigg\}
\end{equation*}
and $F(\cdot)$ is the convex conjugate function of $G( \cdot)$. 

Let $\widetilde G(\omega) = \sum_{i=1}^N \delta_i\,G(\omega_i) q_i^{-1}$ and 
$$\widetilde D_G(\bomega \parallel \bomega^{(0)}) \;=\; \sum_{i=1}^N  \delta_i\, D_G(\omega_i \,\Vert\, \omega_i^{(0)}) \, q_i^{-1}$$ 
be the weighted Bregman divergence generated by $\widetilde G(\omega)$.
The following theorem presents a version of Pythagorean theorem. 

\begin{theorem} 
\label{thm:pythagorean}
Suppose the projection in (\ref{eq:6-3}) is unique. Then, for any element $\bomega \in \mathcal{C}$, we have the equality 
    \begin{equation}
 \widetilde{D}_G ({\bomega} \parallel \bomega^{(0)})  = 
  \widetilde{D}_G({\bomega} \parallel \hat{\bomega}) +  \widetilde{D}_G (\hat{\bomega} \parallel \bomega^{(0)}) , 
  \label{pytha}
  \end{equation}
where $\hat{\bomega}$ is the weights in (\ref{Fwgt2}). 
Moreover, if $\mathcal{C}_1$ is another constraint set with less constraints than $\mathcal{C}$ such that $\mathcal{C}  \subset \mathcal{C}_1$, we have
\begin{equation}
\widetilde D_G(\hat\bomega \parallel \bomega^{(0)}) 
= \widetilde D_G(\bomega_1^\star \parallel  \bomega^{(0)}) + \widetilde D_G(\hat\bomega \parallel  \bomega_1^\star)
\label{pytha2}
\end{equation}
for $\bomega_1^\star = \arg\min_{\omega\in \mathcal{C}_1} \widetilde D_G(\bomega \parallel  \bomega^{(0)})$.
\end{theorem}

Theorem \ref{thm:pythagorean} establishes that the generalized entropy calibration (GEC) solution $\hat\bomega$ is the Bregman projection of the initial weights $\wzero$ onto the set $\mathcal{C}$ that encodes balancing, debiasing, and orthogonality. Consequently, the generalized Pythagorean identity in (\ref{pytha}) 
holds for all $\bomega \in \mathcal{C}$ and so 
$$ \widetilde D_G(\bomega \parallel \bomega^{(0)})  \ge \widetilde D_G(\hat\bomega \parallel  \bomega^{(0)}) $$
holds for all $\bomega \in \mathcal{C}$. Thus, the solution in (\ref{Fwgt2}) indeed minimizes the Bregman divergence from $\bomega^{(0)}$ among the calibration weights in $\mathcal{C}$. 

Also, if $\mathcal{C} \subset \mathcal{C}_1$, the nested identity in (\ref{pytha2})  
quantifies exactly the additional distance paid for imposing the extra constraints in $\mathcal{C}$ that are not in $\mathcal{C}_1$.

\begin{remark} 
The nested identity in (\ref{pytha2}) decomposes the total departure from $\wzero$ into an additive budget:
\[
\underbrace{\tDG(\hat\bomega\Vert\wzero)}_{\text{total}} \,\,\,\,\,\,
=\underbrace{\tDG(\bomega_1^\star\Vert\wzero)}_{\text{baseline constraints in $\mathcal{C}_1$}}
+\;\;\;\; \underbrace{\tDG(\hat\bomega\Vert\bomega_1^\star)}_{\text{price of extras}}.
\]
Thus, additional constraints (like orthogonality) can only increase the divergence from $\wzero$. Monitoring 
$\tDG(\hat\bomega\Vert\bomega_1^\star)$ provides a principled diagnostic for over‑constraint under limited overlap.
\end{remark}

\begin{remark} 
Because $\tDG(\cdot\Vert\cdot)=\sum_i \delta_i q_i^{-1}\DG(\cdot\Vert\cdot)$ is itself a Bregman divergence on the product space, $\hat\bomega$ is the nearest feasible point to $\wzero$ \emph{in the geometry chosen by} $G$ and the metric weights $q_i$. The generator $G$ controls the positivity and extreme values of the weights, while $q(\cdot)$ sets the local metric that can be optimized for efficiency as discussed in Remark 1. 
\end{remark}

\section{High-dimensional covariates}\label{sec:selection-1}

In this section, we extend the proposed method in Section \ref{ch5sec3} to the case of high-dimensional covariates and achieve doubly robust inference under this case. We consider the setting where both the dimensions of the covariates and the basis functions $\bbx_i = (b_1(\bx_i), \ldots, b_p(\bx_i))^{\T}$ for calibration weighting are much larger than the sample size $n = \sum_{i = 1}^{N} \delta_i$. 
When $p > n$, the covariate balancing constraint $\sum_{i=1}^N \delta_i \omega_i \bbx_i = \sum_{i=1}^N \bbx_i$ would not have a solution for $\{\omega_i\}$. Therefore, the proposed calibration weighting method under fixed-dimension settings can not be applied to high-dimension settings. To tackle this problem, we propose a novel soft entropy calibration method in the following. 

Let $\hat{\pi}_i = \pi(\bx_i, \hat{\bphi})$ be a regularized estimation of propensity scores, for example, the penalized maximum likelihood estimation or regularized calibration estimation \citep{tan2020regularized}. Recall that $\bz_i = \bz_i(\hat {\bm \phi})$ denotes the augmented calibration functions, where $\bz_i(\bm \phi) = (\bbx_i^{\T},\partial_\phi
 {g}_i^{\T}(\bphi) q_i^{-1} , g(1 / \pi(\bx_i, \bphi)) q_i^{-1})^{\T}$ where 
$\partial_\phi
 {g}_i^{\T}(\bphi) =g^{\prime}( 1 / \pi(\bx_i, \bphi) ) \{\pi(\bx_i, \bphi)\}^{-2}  \pi^{\prime}(\bx_i, \bphi)^{\T}$.   
From Theorem \ref{thm:1}, the proposed calibration estimator is asymptotically equivalent to a bias-corrected prediction estimator due to the covariate balancing constraints. 
For high-dimensional covariates, as the exact balance is no longer possible for all covariates, we consider obtaining an ``ideal'' calibration direction and impose the exact balancing constraint on this direction. This motivates us to consider the regularized weighted regression 
\begin{equation}
\hat{\bgamma}_{\rm hd} = \argmin_{\bgamma \in \mathbb{R}^{p + p_0 + 1}} \frac{1}{N} \sum_{i=1}^N \delta_i \{g^{\prime}(\hat{\pi}_i^{-1}) q_i^{-1}\}^{-1} ( y_i - \bz_i^{\T} \bgamma )^2 + \tau_1 |\bgamma|_1, 
\label{eq:regression-hd}
\end{equation}
where $\hat{\pi}_i$ is the estimated propensity score, $|\cdot|_1$ denotes the vector $\ell_1$ norm, and the penalty parameter $\tau_1 \to 0$ as $N, p \to \infty$. Let $\bu_i(\bm \phi) = (\bbx_i^{\T}, \partial_\phi
 {g}_i^{\T}(\bphi) q_i^{-1} )^{\T}$ be the sub-vector of $\bz_i(\bm \phi)$ without the last dimension. 
Let $\bu_i = \bu_i(\hat{\bm \phi})$ and $\bar{\bu} = \sum_{i=1}^N \bu_i / N$.
To obtain calibration weights, we solve the constraint optimization problem: 
\begin{eqnarray}
\hat{\bm \omega}_{\rm hd} &=& \argmin_{\omega_i \in \mathcal V}\sum_{i=1}^N \delta_i G( \omega_i) q_i^{-1}, 
\mbox{ \ subject to \ } \sum_{i = 1}^{N} \delta_i \omega_i = N, \label{Wel-HD} \\
&& \sum_{i =1}^N \delta_i \omega_i g(\hat{\pi}_i^{-1}) q_i^{-1}  = \sum_{i=1}^N g(\hat{\pi}_i^{-1}) q_i^{-1} \label{con4-HD} \\
&& \sum_{i=1}^N \delta_i \omega_i \bm z_i^{\T} \hat{\bgamma}_{\rm hd} = \sum_{i=1}^N \bm z_i^{\T} \hat{\bgamma}_{\rm hd} \mbox{ \ and } \label{Calib-HD-exact} \\
&& \frac{1}{N} \bigg| \sum_{i=1}^N \delta_i \omega_i (\bm u_i - \bar{\bm u}) \bigg|_{\infty} \leq \tau_2, \label{Calib-HD} 
\end{eqnarray}
where $\hat{\bm \omega}_{\rm hd} = (\hat{\omega}_{{\rm hd}, i}: \delta_i = 1)$, and $\tau_2$ is a regularization parameter for soft calibration, which diminishes to zero as $N, p \to \infty$. Note that the debiasing calibration constraint in (\ref{con4-HD}) is the same as the one in (\ref{con4}) for the fixed-dimensional case, meaning the function $g(\hat{\pi}_i^{-1})$ for the estimated propensity score is exactly balanced. Compared to the exact covariate balancing constraints in (\ref{Calib}) and (\ref{con-phi}) under the fixed-dimensional setting, we impose the soft covariate balancing constraints in (\ref{Calib-HD}) for high-dimensional covariates. 
However, the projection $\bm z_i^{\T} \hat{\bgamma}_{\rm hd}$ by the estimated direction $\hat{\bgamma}_{\rm hd}$ is exactly balanced in (\ref{Calib-HD-exact}), which we call the projection calibration constraint. This is also related with the model calibration of \cite{wu2001}. 

Let $\tilde{\bu}_i = \bu_i - \bar{\bu}$ and $\blambda_{\rm hd} = (\lambda_{\rm hd, 1}, \lambda_{\rm hd, 2}, \lambda_{\rm hd, 3}, \blambda_{\rm hd, 4}^{\T})^{\T}$, where $\blambda_{\rm hd, 4} \in \mathbb{R}^{p + q}$. Similar to (\ref{dual}), the dual problem of the constraint optimization in (\ref{Wel-HD})--(\ref{Calib-HD}) can be expressed as
\begin{equation}
\begin{split}
\hat{\bm \lambda}_{\rm hd} = \argmin_{\bm \lambda_{\rm hd}} & \frac{1}{N}\sum_{i = 1}^N \delta_i q_i^{-1} F\{(\lambda_{\rm hd, 1} + \lambda_{\rm hd, 2} g(\hat{\pi}_i^{-1}) q_i^{-1} + \lambda_{\rm hd, 3} \bm z_i^{\T} \hat{\bgamma}_{\rm hd} + \blambda_{\rm hd, 4}^{\T} \tilde{\bu}_i) / q_i^{-1}\} \\
& - \lambda_{\rm hd, 1} - \lambda_{\rm hd, 2} \bar{z}_{p + p_0 + 1} - \lambda_{\rm hd, 3} \bar{\bm z}^{\T} \hat{\bgamma}_{\rm hd} + \tau_2 |\bm \lambda_{\rm hd, 4}|_1, 
\end{split}
\label{dual-HD}
\end{equation}
where $\bar{\bz} = (\bar{z}_1, \ldots, \bar{z}_{p + p_0 + 1})^{\T} = \sum_{i = 1}^{N} \bz_i / N$ and $\hat{\bm \lambda}_{\rm hd} = (\hat{\lambda}_{\rm hd, 1}, \hat{\lambda}_{\rm hd, 2}, \hat{\lambda}_{\rm hd, 3}, \hat{\blambda}_{\rm hd, 4}^{\T})^{\T}$. 
Then, the soft calibration weights $\hat{\omega}_{{\rm hd}, i} = \hat{\omega}_{{\rm hd}, i}(\hat{\bm \lambda}_{\rm hd}, \hat{\bphi}, \hat{\bgamma}_{\rm hd})$ from the constrained entropy minimization problem in (\ref{Wel-HD}) satisfy 
\begin{equation}
   \hat{\omega}_{{\rm hd}, i}(\hat{\bm \lambda}_{\rm hd}, \hat{\bphi}, \hat{\bgamma}_{\rm hd}) = f \{ (\hat{\lambda}_{\rm hd, 1} + \hat{\lambda}_{\rm hd, 2} g(\hat{\pi}_i^{-1}) q_i^{-1} + \hat{\lambda}_{\rm hd, 3} \bm z_i^{\T} \hat{\bgamma}_{\rm hd} + \hat{\blambda}_{\rm hd, 4}^{\T} \tilde{\bu}_i) / q_i^{-1} \}
\label{eq:weight-est-HD}
\end{equation}
for the set of observed responses $\{i: \delta_i=1\}$ under high-dimensional covariates. 

The proposed high-dimensional generalized entropy calibration (GEC-HD) estimator of $\theta = \bbE(Y)$ is
\begin{equation}
    \hat{\theta}_{\rm GEC}^{(\rm hd)} = N^{-1} \sum_{i=1}^N \delta_i \hat{\omega}_{{\rm hd}, i} y_i = N^{-1} \sum_{i=1}^N \delta_i \hat{\omega}_{{\rm hd}, i}(\hat{\bm \lambda}_{\rm hd}, \hat{\bphi}, \hat{\bgamma}_{\rm hd}) y_i.
\label{eq:GEC-HD}
\end{equation}
Similar to the arguments in Section \ref{ch5sec3}, let $\blambda_{\rm hd}^*$, $\bm \phi^*$ and $\bgamma_{\rm hd}^*$ be the probability limits of $\hat{\blambda}_{\rm hd}$, $\hat {\bm \phi}$ and $\hat{\bgamma}_{\rm hd}$ under either a correctly specified OR or RP model, respectively. 
In the following, we heuristically explain that the proposed estimator $\hat{\theta}_{\rm GEC}^{(\rm hd)}$ also has the doubly robust inference property under the high-dimensional setting.

Recall that $p$ and $p_0$ are at the same order.
Let $s = \max\{ |\blambda_{\rm hd}^*|_0, |\bm \phi^*|_0, |\bgamma_{\rm hd}^*|_0\}$, where $|\bm a|_0$ denotes the number of nonzero elements in a vector $\bm a$. Under some regularity conditions, following the arguments in \cite{XiaQiu2025}, it could be shown that $|\hat{\bphi} - \bm \phi^*|_1 = O_p\{s\sqrt{\log(p)/n}\}$, $|\hat{\bgamma}_{\rm hd} - \bgamma_{\rm hd}^*|_1 = O_p\{s \sqrt{\log(p)/n}\}$ and $|\hat{\bm \lambda}_{\rm hd} - \blambda_{\rm hd}^*|_1 = O_p\{\sqrt{s^3 \log(p)/n}\}$ by choosing $\tau_1 = c_1\sqrt{\log(p)/n}$ and $\tau_2 = c_2\sqrt{s\log(p)/n}$ for positive constants $c_1$ and $c_2$.
Due to the projection calibration constraint in (\ref{Calib-HD-exact}), we can write $\hat{\theta}_{\rm GEC}^{(\rm hd)}$ as 
\begin{equation}\label{eq:HDexpansion-1}
    \begin{split}
        \hat{\theta}_{\rm GEC}^{\rm hd} &= \frac{1}{N} \sum_{i=1}^N \delta_i \hat{\omega}_{{\rm hd}, i} y_i + \frac{1}{N} \bigg( \sum_{i=1}^N \bz_i - \sum_{i=1}^N \delta_i\hat{\omega}_{{\rm hd}, i} \bz_i \bigg)^{\T} \hat{\bgamma}_{\rm hd} \\
        &= \frac{1}{N} \sum_{i=1}^N \delta_i \hat{\omega}_{{\rm hd}, i} y_i + \frac{1}{N} \bigg( \sum_{i=1}^N \bz_i - \sum_{i=1}^N \delta_i\hat{\omega}_{{\rm hd}, i} \bz_i \bigg)^{\T} \bgamma_{\rm hd}^* + O_p\{ s^{3/2} n^{-1} \log(p) \},
    \end{split}
\end{equation}
where the last equation is due to the soft calibration constraints on $\bu_i$ in (\ref{Calib-HD}). 

If the OR model $m(\bx_i) = \bbx_i^{\T} \bbeta_0$ is correctly specified, we have 
\begin{eqnarray*}
        \hat{\theta}_{\rm GEC}^{\rm hd} &=& \frac{1}{N} \sum_{i=1}^N \bm b_i^{\T} \bbeta_0 + \frac{1}{N} \sum_{i=1}^N \delta_i \hat{\omega}_{{\rm hd}, i} e_i + O_p\{ s^{3/2} n^{-1} \log(p) \} \\ 
        &=& \frac{1}{N} \sum_{i=1}^N \bm b_i^{\T} \bbeta_0 + \frac{1}{N} \sum_{i=1}^N \delta_i \omega_{{\rm hd}, i}^* e_i + O_p\{ s^{3/2} n^{-1} \log(p) \},
\end{eqnarray*}
where $\omega_{{\rm hd}, i}^* = \hat{\omega}_{{\rm hd}, i}(\bm \lambda_{\rm hd}^*, \bphi^*, \bgamma_{\rm hd}^*)$, and the small order term in the last equality above is due to the convergence of $\hat{\blambda}_{\rm hd}$, $\hat {\bm \phi}$ and $\hat{\bgamma}_{\rm hd}$ and the large deviation bound on $N^{-1} \big| \sum_{i = 1}^{N} \tilde{h}(\bx_i) e_i \big|$ for any function $\tilde{h}(\cdot)$ of $\bx_i$. Since $n$ and $N$ are at the same order, $s^{3/2} n^{-1} \log(p) = o(N^{-1/2})$ if $s^3 \log^2(p) = o(N)$, which means the small order term $O_p\{ s^{3/2} n^{-1} \log(p) \}$ can be ignored for the inference procedure of $\hat{\theta}_{\rm GEC}^{\rm hd}$.

On the other hand, if the RP model is correctly specified, we have $\omega_{{\rm hd}, i}^* = \pi_0^{-1}(\bx_i)$ and 
\begin{equation*}
    \begin{split}
        \hat{\theta}_{\rm GEC}^{\rm hd} 
        &= \frac{1}{N} \sum_{i=1}^N \delta_i \pi_0^{-1}(\bx_i) y_i + \frac{1}{N} \bigg( \sum_{i=1}^N \bz_i^* - \sum_{i=1}^N \delta_i \pi_0^{-1}(\bx_i) \bz_i^* \bigg)^{\T} \bgamma_{\rm hd}^* \\
        &+ \frac{1}{N} \sum_{i=1}^N \delta_i \{\hat{\omega}_{{\rm hd}, i} - \pi_0^{-1}(\bx_i)\} (y_i - \bz_i^{\T} \bgamma_{\rm hd}^*) + \frac{1}{N} \sum_{i=1}^N (1 - \delta_i \pi_0^{-1}(\bx_i)) (\bz_i - \bz_i^*)^{\T} \bgamma_{\rm hd}^*
    \end{split}
\end{equation*}
by ignoring the term $O_p\{ s^{3/2} n^{-1} \log(p) \}$ in (\ref{eq:HDexpansion-1}).
Similar to the argument after (\ref{eq:13}), from the KKT condition to the regularized weighted regression in (\ref{eq:regression-hd}), it can be shown that the last two terms in the above equation converge to zero at a rate faster than $O_p(N^{-1/2})$ under suitable conditions. 

Those results imply the Neyman orthogonality of the proposed high-dimensional GEC estimator $\hat{\theta}_{\rm GEC}^{(\rm hd)}$ to the nuisance parameters $\blambda$, $\bphi$ and $\bgamma$, and its doubly robust inference property. 
Therefore, the asymptotic expansion of $\hat{\theta}_{\rm GEC}^{(\rm hd)}$ is
\begin{equation}
\hat{\theta}_{\rm GEC}^{(\rm hd)} = \frac{1}{N} \sum_{i=1}^N \delta_i \omega_{{\rm hd}, i}^* y_i + \frac{1}{N} \sum_{i=1}^N (1 - \delta_i \omega_{{\rm hd}, i}^*) \bz_i^{* \T} \bgamma_{\rm hd}^* + o_p(N^{-1/2})
\label{eq:asy-exp-hd}\end{equation}
under either a correctly specified OR or RP model. 
The confidence interval of $\theta$ can be constructed based on $\hat{\theta}_{\rm GEC}^{(\rm hd)}$ and its linearization result in (\ref{eq:asy-exp-hd}) in the same way as (\ref{eq:Vest}) and (\ref{eq:CI}). 
The asymptotic result in (\ref{eq:asy-exp-hd}) can be rigorously proved following the same procedure as the proof of Theorem \ref{thm:1} under suitable conditions for high-dimensional estimation.
We omit its rigorous proof in this paper and mainly focus on our main methodology contribution of generalized entropy calibration.

\section{Numerical experiments}
\label{ch5sec5}
\subsection{Simulation study}
\label{sim1}



We performed a limited simulation study to investigate the proposed estimators. 
For $i=1,\ldots,N=1{,}000$, $(\bm x_i,Y_i,\delta_i)$ are generated $B=1{,}000$ times repeatedly, 
where $\bm x_i=(1, x_{i1},x_{i2},x_{i3})^\top$. 
The following two outcome regression (OR) models were considered:
\begin{align*}
   \text{O1: }& Y_i \;=\; 1 + x_{i1} - x_{i2} + e_i, \\ 
   \text{O2: }& Y_i \;=\; 1 + x_{i1} - x_{i2} + 0.5\,x_{i1}x_{i2} + 0.3\,(x_{i2}^2-1) + e_i,
\end{align*}
where $X_{ij}\stackrel{\text{i.i.d.}}{\sim}N(2,1)$ for $j=1,2,3$. 
In addition, we considered two variance models for the error distribution:
\begin{align*}
   \text{V1: }& e_i \sim N(0,1), \\
   \text{V2: }& e_i \sim N\!\left(0,\,\max\{0.5,\,x_{i2}^2/4,\,x_{i3}^2/4\}\right).
\end{align*}
Thus, the simulation design follows a $2\times 2$ factorial structure determined by 
the outcome regression model (O1 vs.\ O2) and the variance model (V1 vs.\ V2). 

The sample is obtained by stratified sampling. 
Specifically, the population is divided into four strata according to whether $x_{i2}$ and $x_{i3}$ are above or below their mean value (2).
That is, each unit is assigned to one of the four groups defined by
\((x_{i2} \le 2, x_{i3} \le 2), (x_{i2} \le 2, x_{i3} > 2), (x_{i2} > 2, x_{i3} \le 2), \text{ and } (x_{i2} > 2, x_{i3} > 2).\)
From these four strata, fixed samples of sizes $n_h = (150, 100, 100, 50)$ are drawn without replacement.
The inclusion probability for unit $i$ in stratum $h$ is then 
$\pi_i = n_h/N_h$, where $N_h$ is the population size of stratum $h$. In our setup, we set  $\bm x_{i, {\rm RP}} = (1, x_{i2}, x_{i3})^\top$ and assume $\pi_i = \mbox{expit}(\bm x_{i, {\rm RP}}^\top {\bm \phi})$ as the working model for the response mechanism. 
To estimate the parameter $\bm \phi$ of the working PS model, we use the maximum likelihood estimator. That is, we solve
    $$
    \frac{1}{N}\sum_{i = 1}^N \del{\delta_i - \pi(\bm x_{i, \rm RP}, \bm \phi)} \bm x_{i, {\rm RP}} = 0.
    $$

The parameter of interest is the population mean, $\theta = \mathbb E(Y)$. 
We consider four scenarios depending on whether the outcome regression model 
and the variance model are correctly or incorrectly specified. 
Letting $\bm x_{i, {\rm OR}} = (1, x_{i1}, x_{i2})^\top$, and $\bm x_{i, {\rm RP}} = (1, x_{i2}, x_{i3})^\top$, we compare the following estimators: 
\begin{enumerate}
    \item The Inverse Probability Weighting (IPW) estimator based on the logistic regression model for the response probability $\pi(\bm x_{i, {\rm RP}}, \hat{\bm \phi}) = \mbox{expit}(\bm x_{i, {\rm RP}}^\top \hat{\bm \phi})$. 
    \item The augmented inverse probability weighting (AIPW)   estimator in \eqref{eq:aipwq} using the logistic regression model for the response probability $\pi(\bm x_{i, {\rm RP}}, \hat{\bm \phi}) = \mbox{expit}(\bm x_{i, {\rm RP}}^\top \hat{\bm \phi})$ and the regression coefficient $\hat{\bbeta}_{\rm GLS}(\bm x_{i, {\rm OR}})$ in \eqref{gls} using O1 as the working model. We consider two types of the AIPW estimators: (i) 
 The AIPW estimator in (\ref{eq:aipwq}) with $q_i=1$ (AIPW1), (ii) The AIPW estimator in (\ref{eq:aipwq}) with $q_i=q_i^*$ (AIPW2), where $q_i^* = \hat{\pi}_i^{\hat{\kappa} -1}$ and $\hat{\kappa}$ is the minimizer of $\hat{L}( \kappa)$:
\begin{equation}
\hat{L} (\kappa) = \frac{1}{N}\sum_{i=1}^N \delta_i \bigg\{ \frac{N}{N\hat{\pi}_i} - \hat{\Delta}_b^\top  \hat{\mathbf{M}}_q^{-1}(\kappa) q(\bx_i; \kappa)\, \mathbf{b}_i 
\bigg\}^{2} \tilde{v}_i, 
\end{equation}
with $q(\bx_i; \kappa) = \hat{\pi}_i^{\kappa-1} $ and $\tilde{v}_i = 1$ for V1 and $\tilde{v}_i = (y_i - \bz_i^\top \hat{\bm \gamma})^2$ for V2.      
    \item The proposed generalized entropy calibration (GEC) estimator in \eqref{Wel} using empirical likelihood (EL, \( G(\omega) = -\log \omega \)), exponential tilting (ET, \( G(\omega) = \omega \log \omega - \omega\)), or Hellinger distance entropy (HD, \( G(\omega) = -4 \sqrt{\omega} \)).
\begin{itemize}
    \item GEC1: Uses the calibration constraint \eqref{Calib} on  $\bm x_{i, {\rm OR}}$  
    and the debiasing constraint \eqref{con4} with $q_i=1$.  
      \item GEC2: Uses the calibration constraint \eqref{Calib} on  $\bm x_{i, {\rm OR}}$  
    and the debiasing constraint \eqref{con4} with $q_i = q( \bx_i; \kappa) = \hat{\pi}_i^{\kappa-1}$. The parameter $\kappa$ is chosen by minimizing 
$$ M( \kappa) = \sum_{i = 1}^N \delta_i \big\{ g^{-1} (  q( \bx_i; \kappa) \bz_i^\top  \hat{\bm \lambda} ) \big\}^2 \tilde{v}_i $$
with respect to $\kappa$, where $\hat{\bm \lambda}$ is the solution to \eqref{dual}, and $\tilde{v}_i = 1$ for V1 and $\tilde{v}_i = (y_i - \bz_i^\top \hat{\bm \gamma})^2$ for V2.
 \item GEC3: Uses \eqref{Calib}, \eqref{con4} and the orthogonal calibration constraint \eqref{con-phi} with $q_i=1$.
    \item GEC4: Uses \eqref{Calib}, \eqref{con4} and the orthogonal calibration constraint \eqref{con-phi} with $q_i= \hat{\pi}_i^{\kappa-1}$ using $\kappa$ optimized in GEC2.
\end{itemize}
\end{enumerate}

\begin{table}[!ht]
\centering
\begin{tabular}{@{}lrrrrrrrr@{}}
\toprule
               & \multicolumn{4}{c}{Bias $(\times 10^3)$} & \multicolumn{4}{c}{RMSE $(\times 10^3)$} \\ \cmidrule(l){2-9} 
 & O1V1     & O1V2     & O2V1     & O2V2    & O1V1    & O1V2     & O2V1     & O2V2     \\ \midrule
IPW            & -0.9     & -3.5     & 12.7     & 4.6     & 85.2    & 101.6    & 140.5    & 152.8    \\
AIPW1           & 0.2      & -2.9     & 12.7     & 4.1     & 70.0    & 89.3     & 100.0    & 113.3    \\
AIPW2          & 0.2      & -2.7     & 5.9      & 1.0     & 69.9    & 88.6     & 98.4     & 112.0    \\
\hdashline EL1 & 0.3      & -2.6     & 11.9     & 3.3     & 70.0    & 88.9     & 97.9     & 111.6    \\
EL2            & 0.3      & -2.5     & 9.9      & 3.6     & 69.8    & 87.9     & 97.2     & 111.0    \\
EL3            & 0.6      & -2.4     & 11.6     & 3.3     & 70.2    & 89.2     & 95.8     & 109.4    \\
EL4            & 0.5      & -2.3     & 9.5      & 2.4     & 70.2    & 88.6     & 95.7     & 110.0    \\
\hdashline ET1 & 0.3      & -2.7     & 12.2     & 3.6     & 70.0    & 88.7     & 98.7     & 112.2    \\
ET2            & 0.4      & -2.5     & 9.2      & -2.4    & 69.7    & 86.5     & 97.2     & 109.1    \\
ET3            & 0.4      & -2.5     & 11.8     & 3.0     & 70.1    & 88.5     & 96.4     & 109.6    \\
ET4            & 0.6      & -2.5     & 8.8      & 4.0     & 70.3    & 89.0     & 96.2     & 111.0    \\
\hdashline HD1 & 0.3      & -2.7     & 12.2     & 3.6     & 70.0    & 88.9     & 98.3     & 112.0    \\
HD2            & 0.3      & -2.5     & 10.1     & 2.0     & 69.8    & 87.4     & 97.3     & 110.6    \\
HD3            & 0.3      & -2.6     & 11.9     & 3.2     & 70.1    & 88.5     & 97.9     & 111.1    \\
HD4            & 0.6      & -2.2     & 9.9      & 4.1     & 70.0    & 89.0     & 95.8     & 110.9    \\ \bottomrule
\end{tabular}
\caption{Monte Carlo bias and RMSE ($\times 10^3$) of the estimators under four scenarios defined by the two Outcome Regression models and the two variance models.}
\label{tab:new}
\end{table}

Table \ref{tab:new} presents the Monte Carlo bias and RMSE of the estimators under four combinations of outcome regression and variance models. The IPW estimator performs the worst, showing large bias and RMSE, especially when the outcome regression model is misspecified. Incorporating the outcome regression model reduces bias, as seen in AIPW1 and AIPW2, though their efficiency gains are still limited. By contrast, the GEC estimators (EL, ET, HD) consistently achieve small bias and RMSE across all scenarios. In particular, EL2, ET2, and HD2 yield the lowest RMSE under O1, and adding more calibration constraints—as in EL3 or EL4—further reduces RMSE while keeping bias negligible under O2. Overall, the GEC estimators dominate IPW and perform at least as well as, and often better than, AIPW1. Furthermore, as shown in Table \ref{Table4} of the Supplementary Material, the relative bias of the variance estimators is minor, and the coverage rates of the 95\% confidence interval remain close to the nominal 95\% level across all scenarios. Interestingly, as illustrated in Figure \ref{fig:bd}, for GEC3 estimation, the calibration constraints \eqref{Calib} and \eqref{con4} together accounted for only about 20\% of the total divergence.

\subsection{Real data experiment}
\label{sim2}
We performed another simulation study to investigate the proposed estimators under high-dimensional covariates. We used the 2017-2018 cycle of National Health and Nutrition Examination Survey (NHANES). The NHANES is an ongoing research initiative aimed at evaluating the health and dietary patterns of both adults and children across the United States. To prepare the data for analysis, we first imputed all missing entries using the \texttt{MICE} algorithm \citep{van2011mice}. From the resulting complete dataset with a population size of $N = 9{,}254$, we performed a Monte Carlo simulation of size $B = 500$ to generate repeated samples of artificial missingness. In this study, systolic blood pressure (variable BPXSY1, measured in mmHg) served as the primary outcome variable, denoted by $Y$, while all other variables were used as explanatory covariates with dimension $p = 21$.

To closely replicate the original missingness mechanism, we first fitted a LASSO logistic regression model for $\delta$ and selected the eight most important variables. Using these selected variables, we refitted a standard logistic regression model to estimate the regression coefficients. The estimated coefficients were then used to generate the missingness in the Monte Carlo simulations. The parameter of interest is the mean outcome variable $Y$, estimated as $\mathbb{E}(Y) \approx 121.3$, while the overall missing rate is approximately $\mathbb{E}(\delta) \approx 0.319$.

In this high-dimensional setting, it is unrealistic to assume that the covariates used in the outcome regression and response probability models are known in advance. Therefore, all components of $\bm{x}_i$ are included both when fitting the initial propensity scores and when imposing calibration constraints. The estimators, including IPW and AIPW estimators, in Simulation study 1 are considered, but penalized regression is used to estimate response probability (RP) parameters $\bm{\phi}$ and outcomme regression (OR) coefficients $\bm{\beta}$. For the GEC estimators we used the calibration constraints \eqref{Wel-HD}, \eqref{con4-HD}, \eqref{Calib-HD-exact}, and \eqref{Calib-HD} with $q_i = 1$.
To estimate the RP model parameter $\bm \phi$, we used the regularized maximum likelihood estimation \citep{belloni2014inference}. The hyperparameters were chosen by $5$-fold cross-validation. 

\begin{table}[!ht]
\centering
\begin{tabular}{@{}lcc:ccc@{}}
\toprule
        & IPW     & AIPW    & EL & ET & HD \\ \midrule
Bias    & -0.008  & -0.025  & 0.038     & 0.023    & 0.032    \\
RMSE    & 0.423   & 0.403   & 0.405     & 0.396    & 0.395    \\ \bottomrule
\end{tabular}
\caption{Monte Carlo bias and RMSE of the point estimators.}
\label{Table5}
\end{table}

The simulation results are summarized in Table \ref{Table5}. 
Since the response probability (RP) model is correctly specified, the IPW estimator exhibits a negligible bias.
Similarly, both the AIPW and the proposed GEC estimators yield nearly unbiased estimates with smaller RMSEs. Incorporating the additional calibration constraint \eqref{Calib-HD} in GEC estimators further enhances the efficiency of the GEC estimators compared to the AIPW estimator.


\begin{figure}[htbp]
  \centering
  \begin{minipage}{0.48\linewidth}
    \centering
    \includegraphics[width=\linewidth]{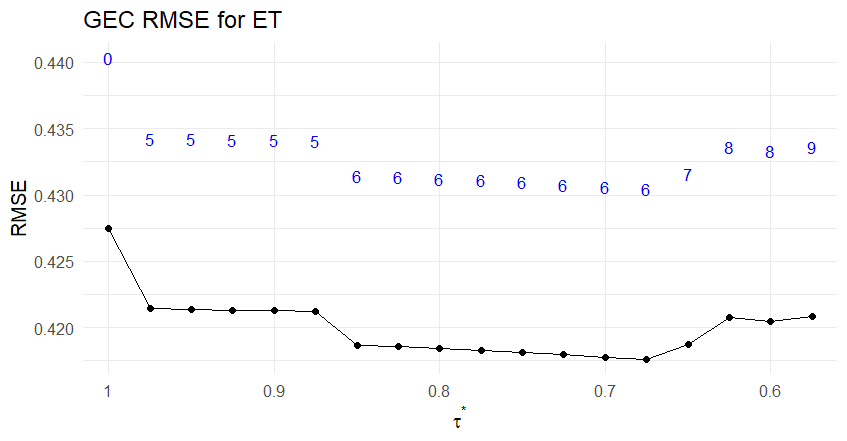}
  \end{minipage}
  \hfill
  \begin{minipage}{0.48\linewidth}
    \centering
    \includegraphics[width=\linewidth]{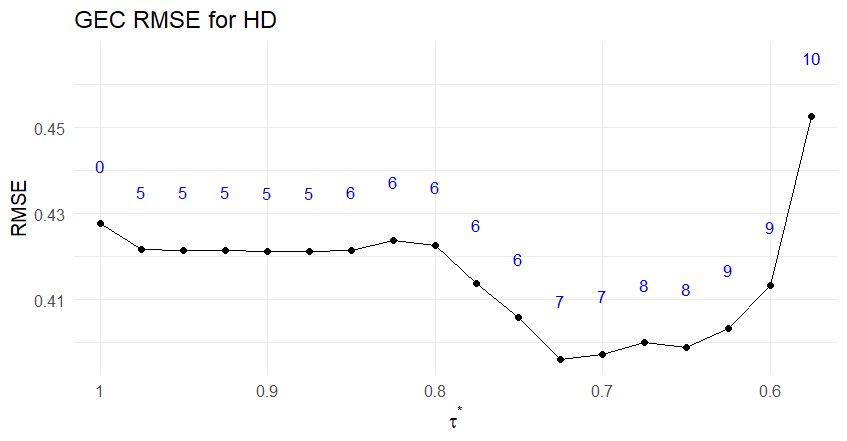}
  \end{minipage}
  \caption{RMSE of the GEC point estimators based on ET (left) and HD (right) entropies across different values of tuning parameter $\tau$. The number of Monte Carlo samples with non-convergent GEC weights is indicated in blue above each point. 
  }
  \label{GEC_RMSE_fig}
\end{figure}

Figure~\ref{GEC_RMSE_fig} presents the sensitivity analysis of the GEC estimators across different values of the tuning parameter $\tau$ used in the calibration constraint \eqref{Calib-HD}. As $\tau$ increases, the RMSE of the estimators tends to decrease up to a certain critical value, after which tighter calibration constraints lead to unstable weights and a larger RMSE. However, larger values of $\tau$ impose stricter calibration constraints, leading to a greater number of Monte Carlo samples with non-convergent GEC weights.



\section{Conclusion}
\label{ch5sec6}

We have proposed an extended class of doubly robust estimator using generalized entropy. The proposed estimator preserves double robustness while introducing desired features. Specifically, it effectively addresses selection bias by integrating a debiasing covariate and also achieves the model-based optimality.  The proposed calibration weights satisfy a version of Pythagorean theorem, which provides a principled diagnostic tool for protecting over-calibration. Soft calibration under high-dimensional covariates is also discussed. 

Future research could explore additional entropy classes, including scaled or shifted entropies being of potential relevance. Multiple propensity scores can be considered by augmenting multiple debiasing constraints \citep{han2013}. Extending these methodologies to cases of missing not at random (MNAR) presents an intriguing challenge. Furthermore, as highlighted by \cite{ma2020}, developing asymptotic theories for scenarios where inclusion probabilities approach zero represents another important area for investigation.

\bibliographystyle{apalike}
\bibliography{ref}

\end{document}